\documentclass[aps,pra,epsf,superscriptaddress,amsmath,amssymb,amsfonts,twocolumn,showpacs,nofootinbib]{revtex4-1}

\usepackage{graphicx}
\usepackage{epsfig}
\usepackage{dcolumn}
\usepackage{bm}
\usepackage{braket}
\usepackage{amsmath}
\usepackage{mathtools}
\usepackage{graphicx,color,xcolor}
\usepackage{hyperref}
\newcommand{\abs}[1]{\left| #1 \right|} 
\usepackage{hyperref}
\usepackage{multirow}
\DeclareMathOperator{\sech}{sech} 

\usepackage[normalem]{ulem} 

\begin{document}
\title{Inducing spin-order with an impurity: phase diagram of the\\ magnetic Bose polaron}
\author{S. I. Mistakidis}
\affiliation{ITAMP, Center for Astrophysics $|$ Harvard \& Smithsonian Cambridge, Massachusetts 02138, USA }
\author{G. M. Koutentakis}
\affiliation{Center for Optical Quantum Technologies, Department of Physics, University of Hamburg, 
Luruper Chaussee 149, 22761 Hamburg Germany} 
\affiliation{The Hamburg Centre for Ultrafast Imaging,
University of Hamburg, Luruper Chaussee 149, 22761 Hamburg, Germany}
\affiliation{IST Austria (Institute of Science and Technology Austria),
Am Campus 1, 3400 Klosterneuburg, Austria}
\author{F. Grusdt}
\affiliation{Department of Physics and Arnold Sommerfeld Center for Theoretical Physics (ASC), 
Ludwig-Maximilians-Universitat M{\"u}nchen, Theresienstr. 37, M{\"u}nchen D-80333, Germany} 
\affiliation{Munich Center for Quantum Science and Technology (MCQST), Schellingstr. 4, D-80799 M{\"u}nchen, Germany}
\author{P. Schmelcher}
\affiliation{Center for Optical Quantum Technologies, Department of Physics, University of Hamburg, 
Luruper Chaussee 149, 22761 Hamburg Germany} \affiliation{The Hamburg Centre for Ultrafast Imaging,
University of Hamburg, Luruper Chaussee 149, 22761 Hamburg, Germany}
\author{H. R. Sadeghpour}
\affiliation{ITAMP, Center for Astrophysics $|$ Harvard \& Smithsonian Cambridge, Massachusetts 02138, USA }

\date{\today}

\begin{abstract} 

We investigate the formation of magnetic Bose polaron, an impurity atom dressed by spin-wave excitations, in a one-dimensional spinor Bose gas. 
In terms of an effective potential model the impurity is strongly confined by the host excitations which can even overcome the impurity-medium repulsion leading to a self-localized quasi-particle state. 
The phase diagram of the attractive and self-bound  repulsive magnetic polaron, repulsive non-magnetic (Fr{\" o}hlich-type) polaron and impurity-medium phase-separation regimes is explored with respect to the Rabi-coupling between the spin components, spin-spin interactions and impurity-medium coupling. 
The residue of such magnetic polarons decreases substantially in both strong attractive and repulsive branches with strong impurity-spin interactions, illustrating significant dressing of the impurity. 
The impurity can be used to probe and maneuver the spin polarization of the magnetic medium while suppressing ferromagnetic spin-spin correlations. 
It is shown that mean-field theory fails as the spinor gas approaches immiscibility since the generated spin-wave excitations are prominent. 
Our findings illustrate that impurities can be utilized to generate controllable spin-spin correlations and magnetic polaron states which can be realized with current cold atom setups.  

\end{abstract}

\maketitle

\section{Introduction} 

Numerous aspects of quasiparticle physics, first introduced by Pekkar and Landau~\cite{landau1933bewegung,landau1948effective}, have been explored in a multitude of ultracold atom experiments~\cite{kohstall2012metastability,koschorreck2012attractive,cetina2016ultrafast,camargo2018creation}.
Their implications range from condensed matter systems~\cite{alexandrov2010advances}, e.g. organic semiconductors~\cite{gershenson2006colloquium}, to chemistry~\cite{bredas1985polarons} and  biophysics~\cite{conwell2005charge}. 
A widely studied quasiparticle in the cold-atom realm is the polaron~\cite{massignan2014polarons,schmidt2018universal}; an impurity dressed by the elementary excitations of the bath with distinct bosonic and fermionic quantum statistics~\cite{yan2020bose,jorgensen2016observation, scazza2017repulsive,cetina2016ultrafast}. 
Fascinating static features of these states include their effective mass~\cite{grusdt2017bose,jager2020strong}, induced interactions~\cite{dehkharghani2018coalescence,petkovic2022mediated} and formation of bound states~\cite{camacho2018bipolarons,will2021polaron,jager2021effect}. 
Typical nonequilibrium phenomena associated with these structures revealed, among others, their induced correlations~\cite{mistakidis2020induced,mistakidis2020many}, dynamical decay and relaxation~\cite{lausch2018prethermalization,mistakidis2020pump,mistakidis2021radiofrequency}, as well as transport~\cite{johnson2011impurity,cai2010interaction,theel2021many}.

In nearly all studies of polaron physics, the bath in which the impurity is immersed in is structureless. 
For the case of a bosonic host, which we are interested in, the amplitude of the phononic excitations triggered by the impurity are suppressed~\cite{mistakidis2019quench,mistakidis2020many}, rendering the observation of the polaron cloud challenging. 
This complication can be alleviated by immersing the impurities into a spinor host where they act as the local perturbers modifying the underlying spin-order. 
These distortions of the magnetic environment give rise to spin-wave excitations~\cite{fukuhara2013quantum} leading to the formation of magnetic Bose polaron. 
This quasiparticle is arguably far less explored than its fermionic counterpart~\cite{kane1989motion,white2002friedel,bohrdt2020dynamical,ji2020dynamical}, and non-magnetic polarons. 
A proposal to detect the polaron cloud, composed
of many-body bound states in the strong-coupling regime, using interferometric techniques in a ferromagnetic gas was made~\cite{ashida2018many}. 
Also, it was argued that a spinor bath can lead to polaronic subdiffusive behavior~\cite{charalambous2020control}.

A spin-carrying medium features the premise of controlling the spin-order and stability of magnetic configurations by deploying an impurity as a probe.
This point of view, we expect to provide a fresh impetus for studying impurity physics and novel many-body phenomena that would be otherwise be challenging to detect. 
Of particular is the stability properties of ferromagnetic domains constituting a longstanding open problem in condensed matter physics \cite{BaberschkeDonath2001}. 
Another promising direction is the understanding of the time-evolution and interactions of magnetic polarons, which is an active topic in spinor fermionic systems~\cite{bohrdt2020dynamical,ji2020dynamical}. They are thought to provide insights into the nature of high-temperature superconductivity~\cite{dagotto1994correlated}. 

An intriguing question is whether a crossover from a non-magnetic to a magnetic Bose polaron can be realized. 
Indeed, since the magnetic character of the medium can be in principle controlled via external radiofrequency fields it might be possible to suppress spin-excitations by appropriately tuning the impurity-medium spin-dependent interactions, in relation to the imposed Rabi-coupling of the spin states.
In fact, it has been demonstrated that the impurity-impurity induced interactions can be tuned via the Rabi-coupling~\cite{compagno2017tunable}. 

To shed light on the impact of an impurity subsystem to a magnetic environment and unravel the magnetic polaron characteristics, we deploy an impurity immersed in a one-dimensional (1D) harmonically trapped spin-$1/2$ Bose gas. 
We use the variational nonperturbative multi-layer multi-configuration time-dependent Hartree method for atomic mixtures (ML-MCTDHX)~\cite{cao2017unified} to calculate the spin-order properties of an impurity immersed in a Bose gas. 
This method treats the impurity-gas system beyond the usual Bogoliubov treatment and thus the Fr{\"o}hlich model~\cite{mistakidis2019dissipative}, as well as to account for boson-boson correlations beyond the Lee-Low-Pines approach~\cite{koutentakis2021pattern}. 

We map out the 1D magnetic polaron phase diagram with respect to the impurity-medium coupling, the spin-spin interactions and the Rabi-coupling of the host spin constituents.
In particular, attractive and self-bound repulsive magnetic polaron states are identified.
The self-bound character of the latter stems from the pronounced magnetic excitations of the host binding the impurity within the gas despite the strong density-density repulsion.
Additionally, the intervals where either the non-magnetic polaron takes place or the quasiparticle picture is no longer valid (due to medium-impurity phase-separation) are delineated. 
The latter two phases are known to also occur in the genuine three-component system~\cite{keiler2021polarons,bighin2021impurity}. 
However, here we demonstrate that it is the presence of Rabi-coupling that proliferates the development of magnetization phenomena and consequently the magnetic polaron~\cite{ashida2018many}.  

The standing spin-wave excitations dressing the impurity are captured by monitoring the spin-flips of each component of the medium. 
The magnetic character of the latter is inferred by the suppression of ferromagnetic spin-spin correlations. 
The latter, being susceptible to parameter variations, are found to be substantial near the miscibility-immiscibility transition of the host spin components.
These manifest at specific interaction-dependent Rabi-couplings and importantly are tunable via the impurity-medium interaction. 
As such the impurity serves as a knob for manipulating the spin-order. 
The magnetic polaron residue decreases with increasing impurity-medium interactions illustrating increased magnetic dressing. Importantly, the residue suppression for large host particle numbers hints towards the susceptibility of the magnetic polaron to the Anderson orthogonality catastrophe~\cite{anderson1967infrared,guenther2021mobile}. 
The energy of the magnetic polaron is negative (positive) for attractive (repulsive) impurity-medium coupling and it increases for larger Rabi-coupling.  

This work is structured as follows. 
Section~\ref{MB_Hamiltonian} introduces the magnetic polaron setting and its intrinsic spin-symmetries while elucidating the role of the impurity to probe spin-wave background excitations. 
The phase diagram of the magnetic polaron is discussed in Section~\ref{phases_magnetic} while its properties are analyzed in detail in Section~\ref{sec:immiscible_bath} including the associated spin-spin correlations and transfer processes. 
A summary of our findings together with future perspectives are provided in Section~\ref{conclusions}. 
Appendix~\ref{app:miscible_bath} presents the persistence of the magnetic polaron for miscible interacting spin components. 
In Appendix~\ref{app:mean_field_binary} we analyze the phase transition induced by the Rabi-coupling in a magnetic Bose gas from a mean-field perspective. In  Appendix~\ref{app:effective_pot} we elaborate on the effective potential approach of the magnetic polaron. Appendix~\ref{app:variational_treat} explicates the variational many-body method used to elucidate the magnetic polaron properties.

\section{Magnetic polaron setup}\label{sec:magnetic_setup} 

To study the magnetic Bose polaron we use a three-component highly particle imbalanced bosonic setting. 
Specifically, a structureless impurity (I) atom is immersed in a spin-$1/2$ bosonic medium having $N=N_{\uparrow}+N_{\downarrow}=100$ particles. 
The multicomponent setup is mass-balanced (unless stated otherwise), namely $m_I=m_{B}$, and all components are confined in the same 1D harmonic trap, $\omega_B = \omega_I = \omega$. 
A corresponding experimental realization is possible by emulating the spin degrees-of-freedom of the medium e.g. via the Rabi-coupled hyperfine states $\ket{\uparrow}\equiv\ket{F=1,m_F=-1}$ and 
$\ket{\downarrow}\equiv\ket{F=2,m_F=1}$ of $^{87}$Rb~\cite{matthews1999watching,williams2000excitation}. 
The impurity atoms might refer to either the $\ket{F=1,m_F=1}$ state of $^{87}$Rb or the isotope $^{85}$Rb.

\subsection{Impurity in a magnetic environment}\label{MB_Hamiltonian}

The underlying many-body Hamiltonian 
\begin{equation}
\begin{split}
\hat{H} =\sum\limits_{a=\uparrow, \downarrow} \hat{H}_a +\sum_{a, a' = \uparrow, \downarrow} &\frac{\hat{H}_{a a'}}{1+\delta_{a a'}}\\
&+\hat{H}_{I}+ 
\sum\limits_{a=\uparrow, \downarrow}\hat{H}_{aI}+\hat{H}_S,
\label{Hamiltonian_total}
\end{split}
\end{equation} 
with the non-interacting parts of the spinor gas $\hat{H}_a=\int dx~\hat{\Psi}^{\dagger}_a(x) \left( -\frac{\hbar^2}{2 m_B} \frac{\partial^2}{\partial x^2}  
+\frac{1}{2} m_B \omega^2 x^2 \right) \hat{\Psi}_a(x)$ where $a=\uparrow, \downarrow$ and the impurity $\hat{H}_{I}=\int dx~\hat{\Psi}^{\dagger}_{I} (x) \left( -\frac{\hbar^2}{2 m_I} \frac{\partial^2}{\partial x^2} +\frac{1}{2} m_I \omega^2 x^2 \right) \hat{\Psi}_{I}(x)$. 
The field operators $\hat{\Psi}_{a} (x)$ act on the spin-$a$ component of the Bose medium and $\hat{\Psi}_{I} (x)$ on the impurity atom. 
Here, we consider ultracold temperatures ensuring the relevance of solely the $s$-wave interactions that are essentially described by a contact potential~\cite{olshanii1998atomic}. 
These are characterized by the spin-spin medium effective couplings $g_{aa'}$ with $a,a'=\uparrow, \downarrow$ and the impurity-spin interaction strengths $g_{Ia}$. 
In this sense, the contact spin-spin interaction term is $\hat{H}_{a a'}=g_{a a'} \int dx~\hat{\Psi}^{\dagger}_{a}(x) \hat{\Psi}^{\dagger}_{a'}(x) 
\hat{\Psi}_{a'} (x)\hat{\Psi}_{a}
(x)$, while the impurity-spin-$a$ interactions are given by $\hat{H}_{Ia}=g_{Ia}\int dx~\hat{\Psi}^{\dagger}_{I}(x) \hat{\Psi}^{\dagger}_{a}(x)\hat{\Psi}_{a}(x)\hat{\Psi}_{I}(x)$. 
\begin{figure*}[ht]
\includegraphics[width=1.0\textwidth]{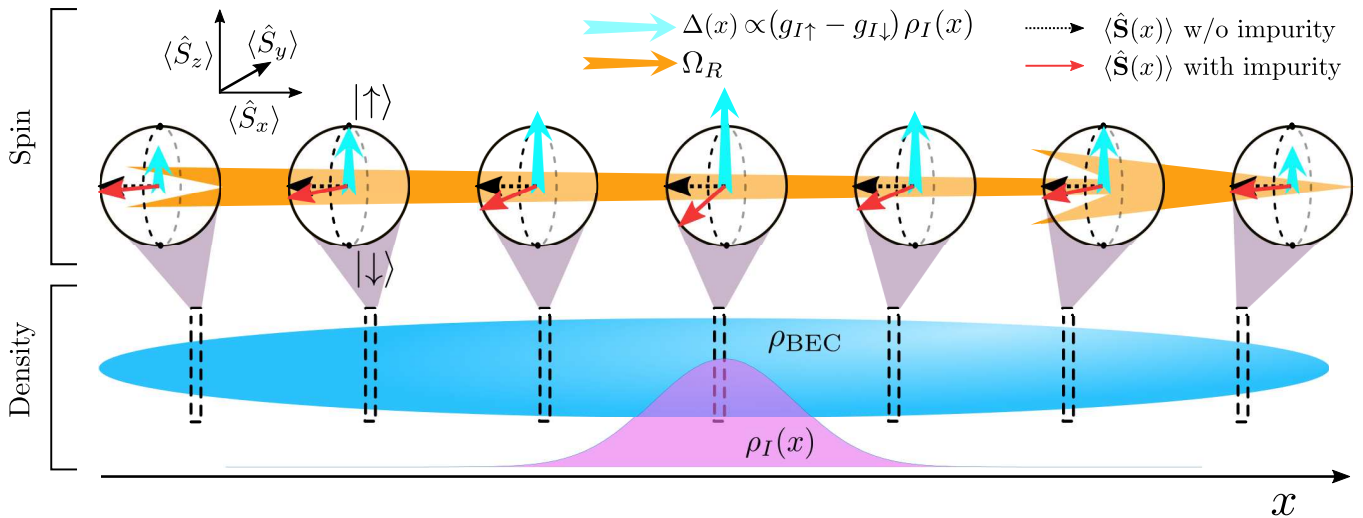}
\caption{Schematic magnetic Bose polaron setup: an impurity with a spatial density profile (purple Gaussian) is immersed in a spin-$1/2$ Bose gas (blue region). Initially, the spins are aligned along the -$x$ direction (black arrows) when a homogeneous radiofrequency $\Omega_R$ (orange long arrow) is applied. The impurity-spin interaction imparts a fictitious magnetic field $\Delta(x)$, modifying the spin polarization (red arrows) and giving rise to  spin-wave excitations thus creating the magnetic Bose polaron.}
\label{fig:schematic_setup} 
\end{figure*} 

These effective coupling constants are related to the three-dimensional $s$-wave scattering length, ${a^s_{\sigma \sigma'}}$ (with $\sigma,\sigma'=\uparrow,\downarrow,I$ being the species index)  and the transversal confinement $\omega_{\perp}$~\cite{olshanii1998atomic}. Consequently, the interaction coefficients can be experimentally adjusted either via Feshbach resonances~\cite{chin2010feshbach,kohler2006production} through ${a^s_{\sigma \sigma'}}$ or utilizing confinement-induced resonances~\cite{olshanii1998atomic} by tuning $\omega_{\perp}$.

The spinor contribution facilitating magnetic processes reads  
\begin{equation}
\hat{H}_S=\frac{\hbar \Omega_{R}}{2} \hat{S}_x + \frac{\hbar \delta}{2} \hat{S}_z. \label{spin_transfer} 
\end{equation}
The Rabi-coupling $\Omega_{R}$ favors the spin-superposition $(1/\sqrt{2})(\ket{\uparrow}-\ket{\downarrow})$ for the medium atoms and therefore admixes different numbers of spin-$\uparrow$ and spin-$\downarrow$ atoms in the ground state. 
The detuning $\delta=\nu -\nu_0$ provides the frequency difference of the Rabi-coupling laser from the corresponding non-interacting ($g_{\uparrow \uparrow}=g_{\downarrow \downarrow}=g_{\uparrow \downarrow}=0$) resonance transition frequency ($\nu_0$) among the $\ket{\uparrow}$ and $\ket{\downarrow}$ states. Here, we consider $\delta=0$, corresponding to a resonant driving in the non-interacting case.

\subsection{Spin-symmetries}\label{sec:spin_symmetries}

The total spin operator is $\hat{\textbf{\textit{S}}}=\int dx~ \hat{\textbf{\textit{S}}}(x)=\int dx \sum_{ab} \hat{\Psi}_a^{\dagger} (x) 
\text{\boldmath$\sigma$}_{ab} \hat{\Psi}_b (x)$, where \text{\boldmath$\sigma$} is the Pauli vector.  
Accordingly, the total spin magnitude operator is expressed as
\begin{equation}
\hat{S}^2 = \hat{S}_+ \hat{S}_- + \hat{S}_z (\hat{S}_z-2),
\label{total_spin}
\end{equation}
where the so-called spin ladder operators $\hat{S}_{\pm}=\hat{S}_x\pm i\hat{S}_y$.
Importantly, the Hamiltonian of Eq.~(\ref{Hamiltonian_total}) preserves the SU(2) symmetry, i.e. $[\hat{S}^2,\hat{H}]=0$ only in the case of $g_{I\uparrow}=g_{I\downarrow}$ and $g_{\uparrow \uparrow} = g_{\downarrow \downarrow} = g_{\uparrow \downarrow}$. 
Otherwise, the eigenvalues of $\hat{S}^2$ are not good quantum numbers. 
Also, in the presence of Rabi-coupling, $\Omega_R \neq 0$, $S_z$ is not conserved, allowing the system become spin-imbalance, i. e. superposition of $\ket{\uparrow}$ and $\ket{\downarrow}$~\footnote{Notice that the term $\propto \delta$ in Eq.~(\ref{spin_transfer}) breaks the $S_x$ symmetry.}.  
Here, we focus on $g_{\uparrow \uparrow}=g_{\downarrow \downarrow}\equiv g$ and therefore in the absence of impurity, the system respects the $\mathbb{Z}_2$ symmetry implying invariance under spin-inversion along the $z$-spin axis. 
However, since the impurity can exhibit different interactions with each of the spin-components, namely $g_{I \uparrow} \neq g_{I \downarrow}$, breaking $\mathbb{Z}_2$ it can thus modify the magnetization of the system because $S_z$ is broken. 

\begin{figure*}[ht]
\includegraphics[width=1.0\textwidth]{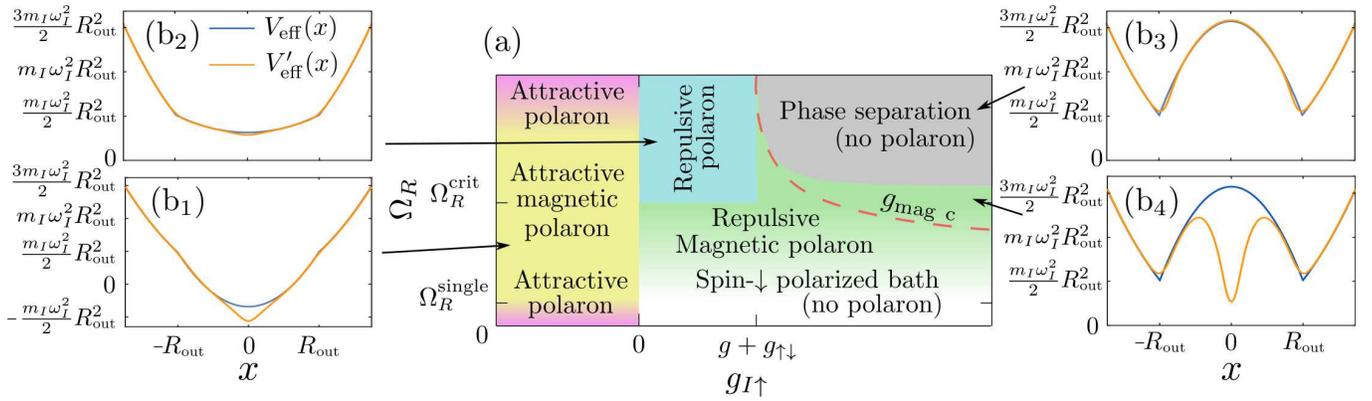}
\caption{(a) Phase diagram of the emergent magnetic polaron states for 
varying impurity-spin ($g_{I \uparrow}$, while we fix $g_{I \downarrow}=0$) interactions and different Rabi-couplings ($\Omega_R$) among the spin-states of the host. Here, $g_{\uparrow \downarrow}^2 > g_{\uparrow \uparrow} g_{\downarrow \downarrow}=g^2$ is further assumed.
The dashed line in (a) indicates $g_{\rm mag~c}$ being the maximal value of $g_{I \uparrow}$ for which the repulsive magnetic polaron is the ground state of the system for a given $\Omega_R$. 
(b$_1$)-(b$_4$) Effective potential experienced by the impurity due to the presence of the spinor gas [see also Eq.~(\ref{gen_eff_pot})]. $\Omega_R^{\rm crit}$ refers to the critical Rabi-coupling for a given $g_{\alpha \alpha'}$, with $\alpha \in \{\uparrow,\downarrow\}$, determining the miscible-immiscible phase-transition of the magnetic Bose gas. $\Omega^{\rm single}_R$ corresponds to the Rabi-coupling below which magnetic effects diminish.
$R_{\rm out}$ and $R_{\rm in}$ denote the TF radii of each spin component and the spin density imbalance respectively.}
\label{fig:schematic_phase} 
\end{figure*}

\subsection{Rescaling and suggested experimental procedure}

In the following, the many-body Hamiltonian described by Eq.~(\ref{Hamiltonian_total}) is expressed in terms of $\hbar \omega$. 
Then, the length, time, Rabi-coupling and interaction strengths are provided with respect to $\sqrt{\hbar/(m \omega)}$, $\omega$, and $\sqrt{(\hbar^3 \omega)/m}$ respectively. 
Experimentally our 1D multicomponent setup is realizable, for instance, by using two hyperfine states of $^{87}$Rb (see above) with $g_{\uparrow \uparrow}=g_{\downarrow \downarrow}= 0.5 \sqrt{\hbar^3 \omega/m}\approx 3.55 \times 10^{-38}$ Jm and a longitudinal (transversal) trap frequency $\omega=2\pi \times 100$ Hz  ($\omega_{\perp}\approx 2\pi \times 5.1$ kHz). 
The respective temperature effects are suppressed as long as the condition $k_BT \ll \frac{3^{4/3}}{16} (\frac{\alpha_{\perp}^2 N_{\uparrow}^2}{a_{\uparrow \uparrow}^s \alpha})^{2/3} \hbar \omega= 316 \hbar \omega \approx 1.5$~$\mu K$ is fulfilled~\cite{pethick2008bose}. 
In the latter expression $k_B$ is the Boltzmann constant, $\alpha_{\perp}=\sqrt{\hbar /(m \omega_{\perp})}$ denotes the transversal confinement length scale and $T$ refers to the temperature of the spinor Bose gas.  

In an experiment our results can be probed by coupling the spin components of the medium via a radiofrequency field~\cite{lavoine2021beyond}. 
The corresponding implementation involves an adiabatic ramp (within a time-interval $\tau \gg \omega_B^{-1}$) of $\Omega_R$. 
The rf field creates a superposition spin state $\ket{\uparrow}$ and $\ket{\downarrow}$ that crucially depends on $g_{a a'}$ and $g_{I a}$.

\subsection{Phases of the spin-$1/2$ Bose gas in the absence of the impurity}

In the absence of impurity, the considered setting reduces to a binary Rabi-coupled Bose-Einstein condensate (BEC)~\cite{abad2013study,tommasini2003bogoliubov,lavoine2021beyond,matthews1999watching}. 
The fundamental building block of such a Rabi-coupled mixture is the case of $\Omega_R=0$ where it boils down to a two-component mixture with particle number conservation in each component. 
Then, it is well-known~\cite{sartori2013dynamics} that the system enters the phase-separated (miscible) state when $g_{\uparrow \downarrow}^2>g_{\uparrow \uparrow} g_{\downarrow \downarrow}$ (in the opposite case). Switching on the Rabi-coupling breaks the $S_z$ spin symmetry of the gas meaning that spin-transfer between the components is allowed. 
As a general rule the effect of the Rabi-coupling reduces the degree of immiscibility (or spin segregation) even when the spin interactions lie deeply inside the immiscible region. 
Specifically, for a larger $\Omega_R$ the gas features a second-order phase transition entering the miscible state and spin-demixing is prohibited~\cite{abad2013study,tommasini2003bogoliubov,matthews1999watching}, see more details in Appendix~\ref{app:mean_field_binary} and also Fig.~\ref{fig:psBEC_pd}.

\subsection{Probing the spin order with the impurity}

The spinor bosonic medium is initially in the ground state configuration, characterized by specific spin-spin interactions $g_{\uparrow \uparrow}=g_{\downarrow \downarrow}\equiv g$ and $g_{\uparrow \downarrow}$ and Rabi-coupling $\Omega_R$. In the absence of the impurity, i.e. $g_{I \uparrow}=g_{I \downarrow}=0$ or equivalently $\langle\hat{H}_{Ia}\rangle=0$, the spins are aligned and the associated spin configuration is almost\footnote{Deviations from the fully polarized (product) state occur due to the development of spin-spin host correlations. Nevertheless, the polarization lies along the $x$ spin axis.} 
identical to a {\it fully} polarized state along the $x$-axis, see Fig.~\ref{fig:schematic_setup}. 
It reads $\ket{P_x}= 2^{-N/2} \bigotimes_{i=1}^{N} \left( \ket{\uparrow}_i-\ket{\downarrow}_i  \right)$ with $i=1,2,\dots,N$. Then, for $g_{I \uparrow} \neq g_{I \downarrow}$ i.e. when the impurity-medium interaction is spin-dependent, the initial polarization vector $\braket{\hat{\bf{S}}}=-N {{\bf e}}_x$, with ${\bf e}_x$ the unit vector, rotates away from the $x$-spin axis. 

Importantly, the localization of the impurity determined by its spatial width acts as a local perturber distorting the polarization of the medium. 
The local spin vector $\braket{\hat{{\bf S}}(x)} \neq \braket{\hat{{\bf S}}(x')}$ for $x\neq x'$ is strongly influenced by the localized impurity dynamics, see Fig.~\ref{fig:schematic_setup}, in turn dressing the impurity with the spin fluctuations; the magnetic polaron. 
An adequate measure to identify the degree of local perturbation in the bath is the spin-spin correlation function (see Eq.~(\ref{spin_corel_func}) below). 
For cases that the spin-order of the host remains intact by the presence of the impurity, the emergent quasiparticle will be referred to as a non-magnetic polaron.

To address the magnetic polaron properties, we employ the variational ML-MCTDHX approach which has been extensively used to study impurity dynamics and spectroscopy~\cite{keiler2021polarons,mistakidis2020induced,mistakidis2020many,mistakidis2021radiofrequency,mistakidis2020pump}. 
ML-MCTDHX is based on the expansion of the many-body wave function in terms of a time-dependent and variationally optimized basis set, see Appendix~\ref{app:variational_treat} for details. 
It is tailored to capture interparticle spatial and spin-spin correlations~\cite{mistakidis2021radiofrequency}, while efficiently truncating the Hilbert, even for mesoscopic particle numbers. 
To expose the role of correlations, we compare our results with the predictions of the three-component Gross-Pitaevskii equation (GPE) [Eq.~(\ref{GrossPitaevskii_3C})]. 
An effective potential model is constructed, see in particular Appendix~\ref{app:effective_pot} and Eq.~(\ref{gen_eff_pot}), for the interpretation of the mechanisms accompanying the generation and properties of the magnetic polaron.

\section{Phase diagram of the magnetic Polaron}\label{phases_magnetic} 

The phase diagram in Fig.~\ref{fig:schematic_phase} is calculated with the variational machinery of ML-MCTDHX. The interpretation of the different phase properties is greatly elucidated with an effective potential [see also Appendix~\ref{app:effective_pot}]. 
By ignoring the elementary BEC excitations, i.e. assuming $\hat{\Psi}_{\alpha}(x)=\hat{\Psi}^{\dagger}_{\alpha}(x)=\sqrt{\rho_{\alpha}(x)}$, the effective potential of an impurity embedded in a spin-$1/2$ Bose gas can be derived as,
\begin{equation}
    V_{{\rm eff}}(x)= \frac{1}{2} m_I \omega_I^2 x^2+ g_{I \uparrow} \rho_{\uparrow} (x) + g_{I \downarrow} \rho_{\downarrow} (x).\label{gen_eff_pot}
\end{equation}
The density of the $a:\{\uparrow,\downarrow\}$ spin components is $\rho_a(x)$, see Eq.~(\ref{eq:reduced1b_den}). 
Typical effective potential approaches \cite{mistakidis2019quench,mistakidis2020many} incorporate the density $\rho_a(x)$ in the absence of impurity ($g_{I \uparrow}=g_{I \downarrow}=0$) thereby neglecting all impurity-bath correlations. 
Here in order to account for the impurity backaction on the medium\footnote{As we will argue below the dominant contribution of impurity-medium correlations stems from the coupling of the spin and spatial degrees-of-freedom in the many-body wave function, see section~\ref{spin_corel}. Spatial medium correlations are suppressed in our setup for weak boson-boson interactions.}, 
we construct an effective potential,  $V'_{\rm eff} (x)$, where $\rho_a(x)$ corresponds now to the spin density obtained from the variational ML-MCTDHX approach. 
This comparison reveals the alteration of the quasiparticle phase diagram due to the above-mentioned  correlations.     

For simplicity, below, we focus on the case where $g_{I \uparrow}$ is finite, but $g_{I \downarrow}=0$. Initially we aim to analyze the competition between the Rabi-coupling $\Omega_R$ and the impurity-spin-$\uparrow$ interaction as shown in Fig.~\ref{fig:schematic_phase} (a). 
We are particularly interested in the case $g_{\uparrow \downarrow} > \sqrt{g_{\uparrow \uparrow} g_{\downarrow \downarrow}}=g$, where spin-excitations dominate [see also Appendix~\ref{app:effective_pot}]. 

To elucidate the main features of the different spin-orders that can emergent in the host we first briefly analyze its phases in the absence of the impurity. 
In the Gross-Pitaevskii mean-field description for the binary Rabi-coupled bosonic bath~\cite{abad2013study}, and within the Thomas-Fermi (TF) approximation [see Appendix~\ref{app:mean_field_binary}], it can be proved that there are three different
$\Omega_R$ regimes with distinct spatial configurations; this result agrees with the correlated ML-MCTDHX calculations (not shown for brevity). 
The first two regimes are determined by the Rabi-coupling in comparison to the effective spin-spin interaction strength. 
In particular, there exists an immiscibility threshold at 
\begin{equation}
\Omega_R^{\rm crit} = n(0)(g_{\uparrow \downarrow}-g),\label{Rabi_threshold}   
\end{equation}
below (above) which the host components are immiscible (miscible) for $g_{I \uparrow}=g_{I \downarrow} =0$. Notice that $n(0)=\rho_{\uparrow} (0)+\rho_{\downarrow} (0)$. 
Importantly, for this choice of interaction parameters, i.e. $g_{\uparrow \downarrow} > g$, the host becomes fully-polarized along the $z$-spin direction for $\Omega_R =0$~\cite{eto2016nonequilibrium,tojo2010controlling,mistakidis2018correlation}. 
Hence, the third $\Omega_R$ regime appears characterized by negligible population of one BEC component. 
To estimate the extent of this regime, we define $\Omega^{\rm single}_R$ below which the minority medium spin-component is populated by less than one atom [see also Appendix~\ref{app:effective_pot}]. 

Next we turn into the coupled impurity-spin-$1/2$ gas setting. 
Let us first consider the case $\Omega_R \gg \Omega_R^{\rm crit}$ where due to the strong effective magnetic field associated with $\Omega_R$, see Eq.~(\ref{spin_transfer}), the host becomes fully-polarized along the $x$ spin-axis. 
As such the BEC components are miscible implying that spin-excitations are rather inert, as discussed in Ref.~\cite{abad2013study}. 
Therefore, only very weak impurity-medium correlations can be induced.
Then the medium can be viewed as an effective single-component BEC with a renormalized coupling $g_{BB}^{\rm eff} = g + g_{\uparrow \downarrow}$ (see also Appendix~\ref{app:effective_pot}).
As a consequence, the Bose polaron formation reduces to the well-understood case of a single-component host~\cite{rath2013field}, where only phononic dressing is possible.
Here, attractive (repulsive) Bose polaron states emerge for $g_{I \uparrow} <0$ ($0 < g_{I \uparrow} < g_{BB}^{\rm eff}$), see Fig.~\ref{fig:schematic_phase}(a), while for $g_{I \uparrow} > g_{BB}^{\rm eff}$, impurity-medium phase separation is observed  
and the polaron ceases to exist~\cite{dehkharghani2018coalescence,mistakidis2020many}.
The latter regime is associated with the temporal orthogonality catastrophe phenomenon elucidated in the dynamical studies in Refs.~\cite{mistakidis2019quench,mistakidis2021radiofrequency}. The effective potential, $V_{\rm eff}(x)$, deforms from a harmonic oscillator potential, with an effective frequency $\omega_{\rm eff} = \omega \sqrt{1 - g_{I \uparrow}/g_{BB}^{\rm eff}}$, for $g_{I\uparrow} < g_{BB}^{\rm eff}$ to a double-well profile in the opposite case, see Fig.~\ref{fig:schematic_phase}(b$_2$), (b$_3$). 
As expected due to the suppressed impurity-medium correlations, $V'_{\rm eff}(x) \approx V_{\rm eff}(x)$ holds.

\begin{figure*}[ht]
\includegraphics[width=1.0\textwidth]{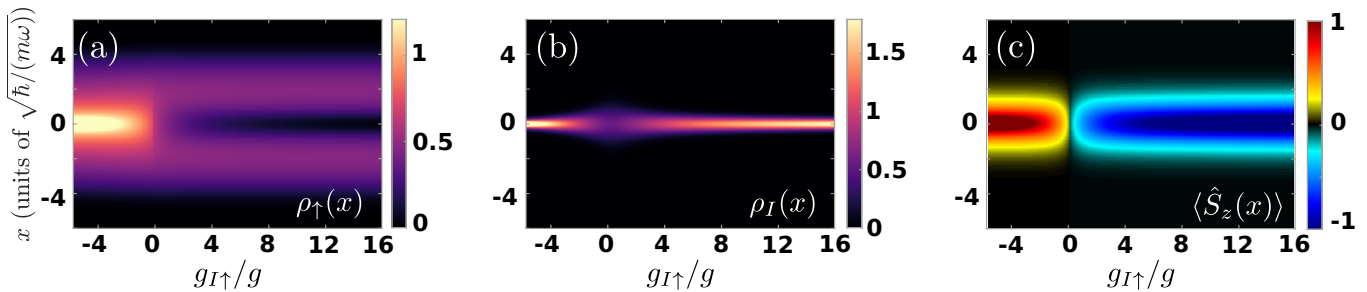}
\caption{Density distributions of an impurity embedded in an immiscible spinor bosonic medium with $\Omega_R/\omega=5$ and $g_{\uparrow \downarrow}/g=1.6$. 
The ground state configurations of (a) the spin-$\uparrow$ and (b) the impurity are depicted for different impurity-spin-$\uparrow$ interactions. 
The density of the spin-$\downarrow$ state is complementary to the spin-$\uparrow$ and it is not shown. 
(c) The non-zero magnetization $\braket{\hat{S}_z(x)}=\rho_{\uparrow}(x)-\rho_{\downarrow}(x)$ of the medium justifies the emergence of standing spin-waves and thus the emergence of the magnetic Bose polaron. 
The atoms of the medium possess spin-spin interactions $g_{\uparrow \uparrow}=g_{\downarrow \downarrow}\equiv g=0.5$, $g_{I \downarrow}=0$ while the system is harmonically trapped.}
\label{fig:1bden_immiscible} 
\end{figure*}

As $\Omega_R \rightarrow \Omega_R^{crit}$, the magnetic degrees-of-freedom of the host gas become relevant. Here new features in addition appear due to spin-excitations. 
In particular, in the vicinity of the transition point between the repulsive Bose polaron and the phase-separated states at $g_{I \uparrow} \gtrsim g + g_{\uparrow \downarrow}$, another interspecies interaction regime appears.
Here, the induced magnetic excitations modify the effective confinement of the impurity. 

This effect is readily observed in Fig.~\ref{fig:schematic_phase}(b$_4$), where $V'_{\rm eff}(x)$ possesses a third well located at $x = 0$, in contrast to the double-well structure of $V_{\rm eff}(x)$. 
This demonstrates the emergence of a magnetic Bose polaron state that is self-localized in the sense that the impurity is confined within the magnetic excitations it induces to its host.
As $\Omega_R \to \Omega_R^{{\rm crit}}$ the magnetic polaron is stable for larger $g_{I \uparrow}$, see Fig.~\ref{fig:schematic_phase}(a).
Due to the interaction energy cost associated with the involved magnetic dressing cloud, this polaron is the ground state of the system\footnote{Notice that the magnetic Bose polaron can be also stable for $g_{I \uparrow} > g_{\rm mag~c}$ despite possessing larger energy than the phase separated ground state. The coexistence of two stable phases within the same interaction interval indicates a hysteresis phenomenon with respect to $\Omega_R$ referring to the presence or absence of the magnetic polaron.} only for $g_{I \uparrow} < g_{\rm mag~c}$, see the dashed line in Fig.~\ref{fig:schematic_phase}(a).
For $g_{I \downarrow} < 0$ the impurity also disturbs the polarization of its host giving rise to magnetic excitations inducing an additional effective attractive force, captured within $V'_{\rm eff}(x)$, which lies beyond the $V_{\rm eff}(x)$ picture, see Fig.~\ref{fig:schematic_phase}(b$_1$). 

For $\Omega_R \leq \Omega_R^{\rm crit}$, the host exhibits a phase separated character around $x \approx 0$, leading to an attractive $V_{\rm eff}(x)$ confining the impurity within the BEC independently of $g_{I \uparrow}$.
Furthermore, the population of the spin-$\uparrow$ component, which interacts with the impurity, is amplified (diminished) for $g_{I \uparrow} < 0$ ($g_{I \uparrow} > 0$) which aids the development of impurity-medium  correlations.
The pronounced magnetic excitations of the host, captured by $V'_{\rm eff}(x)$, 
modify the effective confinement of the impurity $V_{\rm eff}(x)$ and thus the polaron possesses a magnetic character.
The latter ceases to exist for $\Omega_R < \Omega_R^{\rm single}$ since the population of the minority spin-component becomes negligible.
In particular, for repulsive $g_{I \uparrow}>0$ it is energetically preferable for the host atoms to occupy the non-interacting with the impurity spin-$\downarrow$ component, thus preventing quasi-particle formation.
However, $g_{I \downarrow} < 0$, favors the spin-$\uparrow$ configuration and as a consequence an attractive Bose polaron forms, akin to the single-component case analyzed e.g. in Ref.~\cite{mistakidis2020many,mistakidis2020pump,mistakidis2021radiofrequency}.

Finally, we remark that for $g_{\uparrow \downarrow} < g$, no phase-separated regime for the host is encountered, see Appendix~\ref{app:mean_field_binary}.
In this case the phase diagram of the system is similar to the one analyzed above but importantly $\Omega_R^{\rm crit} = 0$ preventing the occurrence of the regimes with $\Omega_R < \Omega_R^{\rm crit}$.
In addition, the repulsive magnetic Bose polaron regime is exhibited within a smaller range of $g_{I \uparrow}$ values, but nevertheless it possesses similar properties to those analyzed above.

\section{Impurity immersed in an immiscible spinor Bose gas}\label{sec:immiscible_bath} 

Since the spin-demixing processes in the case of a spin-$1/2$ Bose gas are enhanced for immiscible spin-spin interactions ($g_{\uparrow \downarrow}^2>g_{\uparrow \uparrow} g_{\downarrow \downarrow}$) our main focus is placed in this interaction regime for providing a clean manifestation of the magnetic polaron. 
As we argue in Appendix~\ref{app:miscible_bath} where we analyze the scenario of miscible spin interactions the spin-demixing is indeed reduced in this regime also in the presence of the impurity. 
Below, we seek the ground state of the composite system. 
Specifically, we choose as a representative setting a spinor medium experiencing $g=0.5$ and $g_{\uparrow \downarrow}/g=1.6$ at a fixed $\Omega_R$ while $g_{I \downarrow}=0$ and $g_{I \uparrow}$ is either attractive or repulsive in order to realize the different magnetic polaron branches. 
\begin{figure*}[ht]
\includegraphics[width=1.0\textwidth]{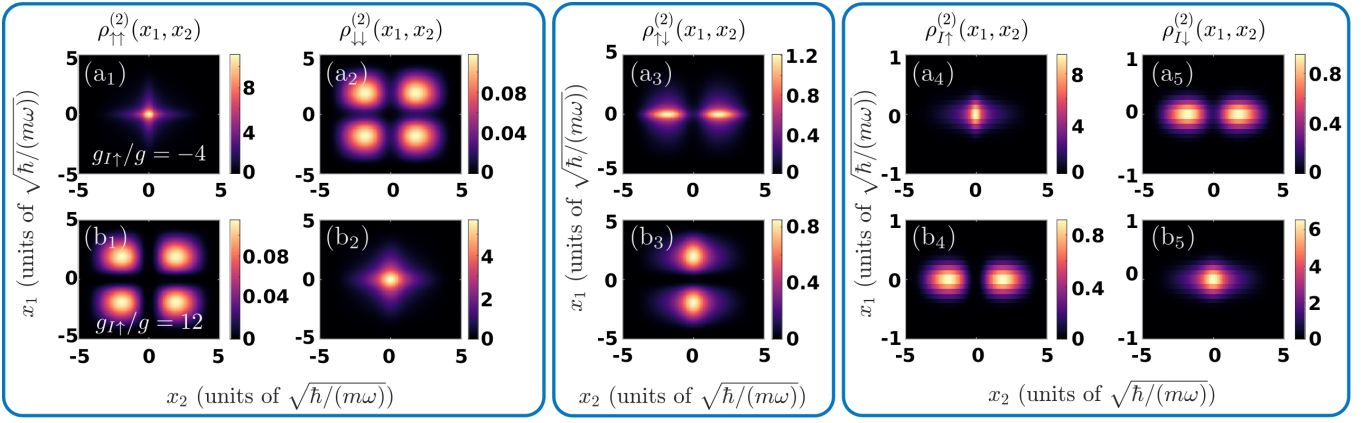}
\caption{Two-body spatial configurations of the spins as well as of the impurity and a spin state for different interactions $g_{I \uparrow}$ (see legends). 
The structural two-body deformations for $g_{I \uparrow}\neq 0$ evince the impact of the impurity on the spin-order (panels (a1)-(a3), (b1)-(b3)) and reveal induced effectively attractive (repulsive) impurity-spin-$\downarrow$ interactions when $g_{I \uparrow}>0$ ($g_{I \uparrow}<0$) (panels (a5), (b5)). 
In the case of $g_{I \uparrow}=0$ the two-body distributions are symmetric by means of $x_1 \leftrightarrow x_2$ and are localized around $x_1 = x_2 =0$ implying that effective interactions vanish (not shown). 
An impurity is immersed in an immiscible spinor medium characterized by $\Omega_R/ \omega=5$, while other system parameters are the same as in Fig.~\ref{fig:1bden_immiscible}.}
\label{fig:2bden_immiscible} 
\end{figure*}

\subsection{Magnetic Polaron configurations and magnetization}\label{distributions_immiscible}

To elucidate the impact of the impurity on the spatial distribution of the medium's spin components we invoke the one-body density of each component 
\begin{equation}
\rho_\sigma(x)=\langle\Psi|\hat{\Psi}_{\sigma}^{\dagger}(x)\hat{\Psi}_{\sigma}(x) |\Psi\rangle.\label{eq:reduced1b_den}
\end{equation}
Here, $\hat{\Psi}_{\sigma}(x)$ is the $\sigma=\uparrow, \downarrow,I$-component bosonic field operator acting 
at position $x$ and $\ket{\Psi}$ denotes the many-body ground state of the three-component system. 
This particle density can be observed by single-shot averaging~\cite{bloch2008many}. 
Notice that $\rho_{\uparrow}(x)+\rho_{\downarrow}(x)=\rho_{TF}(x)=(2/(g+g_{\uparrow \downarrow}))(\mu_B-m\omega^2x^2+\abs{\Omega_R}/2)$, see also Appendix~\ref{app:mean_field_binary}, and thus below we only present the $\rho_{\uparrow}(x)$ since the structures emerging in $\rho_{\downarrow}(x)$ are complementary to it. 
The dependence of $\rho_\sigma(x)$ with the bath-impurity coupling is shown in Fig.~\ref{fig:1bden_immiscible} for $g_{I \uparrow}$ ranging from attractive to strong repulsive with $g_{I\downarrow}=0$ and $\Omega_R=5\omega$. 
This value of $\Omega_R>\Omega_R^{\rm crit} \approx 4.8 \omega$ enforces the miscibility among the spin components in the absence of the impurity, i.e. when $g_{I \uparrow}=0$ or equivalently in the case of no polaron, see also Sec.~\ref{phases_magnetic} and Appendix~\ref{app:mean_field_binary}. 

Finite repulsive impurity-spin-$\uparrow$ interactions lead to a depletion of  $\rho_{\uparrow}(x)$ around the trap center, see Fig.~\ref{fig:1bden_immiscible} (a).
Simultaneously, the impurity features a progressive localization tendency for larger $g_{I \uparrow}$ [Fig.~\ref{fig:1bden_immiscible}(b)] thus forming a quasiparticle. 
This is the first key difference to the non-magnetic repulsive Bose polaron~\cite{mistakidis2019quench,mistakidis2020many} which is known to delocalize for increasingly repulsive impurity-host interactions causing its decay. 
Turning to $g_{I \uparrow}<0$, a behavior similar to the attractive non-magnetic Bose polaron~\cite{mistakidis2020many} is detected.
Here, the spin-$\uparrow$ component shows a sizable density peak at the location of the impurity [Fig.~\ref{fig:1bden_immiscible}(a)], with the concomitant localization of the impurity [Fig.~\ref{fig:1bden_immiscible}(b)]. 
However, the most striking distinction of the quasiparticle realized herein from the non-magnetic polaron stems from the response of the spin-$\downarrow$ state.
Indeed, $\rho_\downarrow(x)$ is complementary to $\rho_\uparrow(x)$, with the former being accumulated (depleted) in the spatial extent of the impurity for repulsive (attractive) $g_{I \uparrow}$ (not shown). This behavior of $\rho_{\downarrow}(x)$ is mediated by the immiscible $g_{\uparrow \downarrow}$ interactions which lead to an effective attraction, $g_{I \downarrow}^{\rm eff}<0$ (repulsion, $g_{I \downarrow}^{\rm eff}>0$), among the spin-$\downarrow$ and the impurity for $g_{I \uparrow} > 0$ ($g_{I \uparrow} < 0$), see also Fig.~\ref{fig:2bden_immiscible} (a$_5$), (c$_5$) and the related discussion in Sec.~\ref{2b_configurations}.
In general, these effective interactions are mediated by the magnonic excitations due finite difference ($g_{I \uparrow} - g_{I \downarrow}$) interaction.
Notably, our observations necessitate the construction of effective Hamiltonians similar to~\cite{mistakidis2019effective} in order to appreciate the strength of induced interactions in three-component settings. 

The above-described response of $\rho_\uparrow(x)$ and $\rho_\downarrow(x)$ hints towards the presence of magnetic processes in the system.
These processes can be analyzed by studying the local magnetization response, $\braket{\hat{S}_z(x)}=\rho_{\uparrow}(x)-\rho_{\downarrow}(x)$ provided in Fig.~\ref{fig:1bden_immiscible}(c).
The emergence of local magnetization in the host is related to a spatially varying effective magnetic field, $\Delta(x) \sim (g_{I \uparrow} - g_{I \downarrow} ) \rho_I(x)$, due to the impurity-spin-$\uparrow$ interactions [see also Fig.~\ref{fig:schematic_setup}] which in turn gives rise to standing spin-waves.
Without the impurity ($g_{I \uparrow}=0$) it holds that $\braket{\hat{S}_z(x)}=0$, due to the polarization of the host in the spin-$x$ axis, while in the case of $g_{I \uparrow}>0$ ($g_{I \downarrow}<0$) we find $\braket{\hat{S}_z(x)}<0$ ($\braket{\hat{S}_z(x)}>0$). It is therefore the presence of the impurity that triggers a $g_{I \uparrow}$-dependent correlated standing spin-wave, which we analyze further in Sec.~\ref{dressing_cloud} and \ref{spin_corel}.    
\begin{figure*}[ht]
\includegraphics[width=1.0\textwidth]{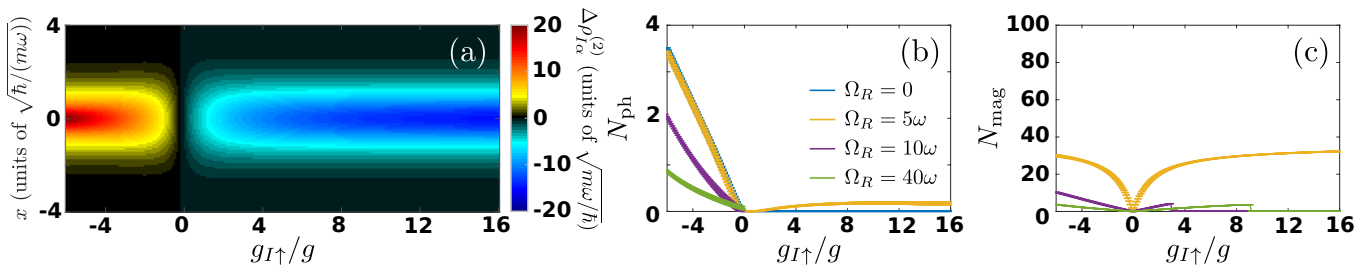}
\caption{(a) Difference between the two-body impurity-medium correlation functions in the comoving impurity frame $\Delta \bar{\rho}^{(2)}_{I a}(x_r)$ for $\Omega_R/\omega=5$. 
The magnetic polaron configuration has a form $\sim \sech^2(x/w)$. 
Dressing cloud of the (b) phononic and (c) magnonic impurity cloud. 
Apparently, the magnonic dressing is stronger justifying the magnetic nature of the polaron.}
\label{fig:dressing} 
\end{figure*}

\subsection{Two-body magnetic polaron distributions}\label{2b_configurations}

A more concrete demonstration of the spatial distributions of the magnetic polaron is obtained  by inspecting the underlying two-body configurations. 
For this reason we determine the diagonal elements of the two-body reduced density matrix 
\begin{equation}
\rho^{(2)}_{\sigma \sigma'}(x_1,x_2)=\langle\Psi|\hat{\Psi}_{\sigma}^{\dagger}(x_2)\hat{\Psi}^{\dagger}_{\sigma'}(x_1)\hat{\Psi}_{\sigma'}(x_1)\hat{\Psi}_{\sigma}(x_2)|\Psi\rangle. \label{2b_reduced}
\end{equation} 
This provides the probability to simultaneously detect a $\sigma$-component boson at position $x_1$ and a $\sigma'$ atom at $x_2$. 
The spatially resolved two-body configurations for the immiscible interacting medium with $\Omega_R/\omega=5$ exhibit involved structures for both attractive [Fig.~\ref{fig:2bden_immiscible} (a$_1$)-(a$_5$)] and repulsive [Fig.~\ref{fig:2bden_immiscible} (b$_1$)-(b$_5$)] impurity-spin-$\uparrow$ interaction strengths. 
Otherwise, for $g_{I \uparrow}=0$ they feature a symmetry under the exchange of $x_1 \leftrightarrow x_2$ and a localization around $x_1 = x_2 =0$. 
This implies that the detection of the two atoms of the same or different components are largely independent, while the atoms are likely to reside around the trap center. 

As explained above, an attractive impurity-medium coupling enforces the spin-$\uparrow$ bosons to lie in the vicinity of the impurity. 
Consequently, $\rho^{(2)}_{\uparrow \uparrow}(x_1,x_2)$ is highly localized [Fig.~\ref{fig:2bden_immiscible} (a$_1$)] around the trap center demonstrating the portion of spin-$\uparrow$ atoms bound to the impurity, see in particular $\rho^{(2)}_{\uparrow \uparrow}(x_1,x_2)$ [Fig.~\ref{fig:2bden_immiscible} (a$_4$)]. 
The low density tails of $\rho^{(2)}_{\uparrow \uparrow}(x_1,x_2)$ infer the existence of spin-$\uparrow$ bosons that remain unbound. 
On the other hand, $\rho^{(2)}_{\downarrow \downarrow}(x_1,x_2)$ exhibits localization in four disjoint spatial domains [Fig.~\ref{fig:2bden_immiscible} (a$_2$)] stemming from the phase-separation between the spin components which is evident by the elongated two-hump shape of $\rho^{(2)}_{\uparrow \downarrow}(x_1,x_2)$ [Fig.~\ref{fig:2bden_immiscible}(a$_3$)]. As a consequence, the impurity and the spin-$\downarrow$ are also phase-separated, despite being non-interacting $g_{I \downarrow} = 0$ [see Fig.~\ref{fig:1bden_immiscible}(b)]. 
Therefore, $\rho^{(2)}_{I \downarrow}(x_1,x_2)$ shows a finite probability at two distinct domains characterized by $|x_\downarrow| > |x_I|$
[Fig.~\ref{fig:2bden_immiscible}(a$_5$)] which certifies the emergence of repulsive induced impurity-spin-$\downarrow$ interactions caused by the attractive $g_{I \uparrow}$.  

Turning to strong repulsive $g_{I \uparrow}$, an inversion in the roles of the spin-$\uparrow$ and spin-$\downarrow$ bosons takes place, compare e.g. Fig.~\ref{fig:2bden_immiscible}(a$_2$) and (b$_1$).
Indeed, in this case the spin-$\uparrow$ atoms and the impurity tend to phase-separate [Fig.~\ref{fig:2bden_immiscible}(b$_4$)] due to their strong repulsion $g_{I \uparrow} \gg g$, while the spin-$\downarrow$ particles are effectively attracted towards the impurity [Fig.~\ref{fig:2bden_immiscible}(b$_5$)], i.e. $g_{I \downarrow}^{\rm eff} < 0$ besides the fact that $g_{I \downarrow} = 0$.
Similar to the non-magnetic polaron case~\cite{mistakidis2019quench,mistakidis2020many}, we observe that also here the quasiparticle character in the repulsive branch is less pronounced than in its attractive counterpart. 
Hence, the emergent patterns in Fig.~\ref{fig:2bden_immiscible}(b$_i$) are less prominent than those depicted in Fig.~\ref{fig:2bden_immiscible}(a$_i$) with $i=1,\dots,5$.

\begin{figure*}[ht]
\includegraphics[width=1.0\textwidth]{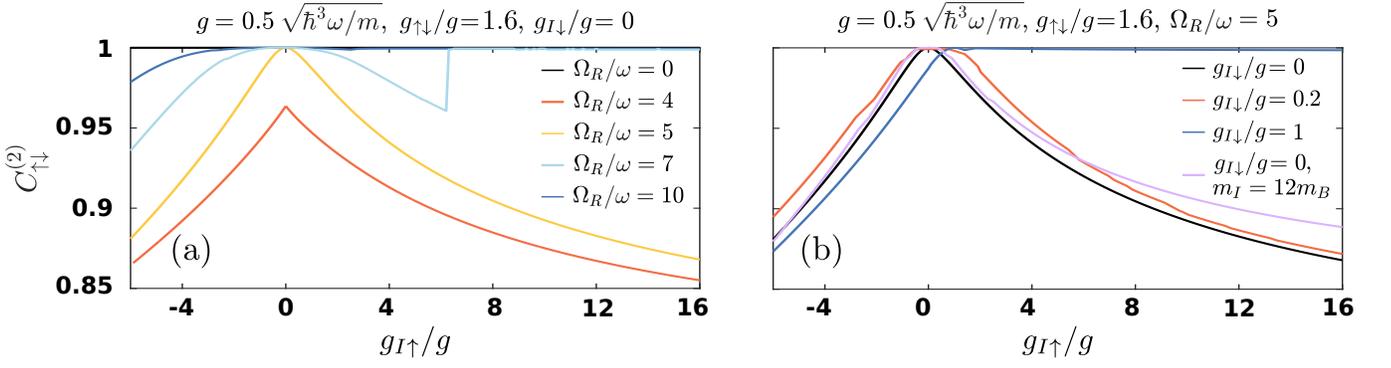}
\caption{Spin-Spin correlation function of an immiscible interacting medium with respect to $g_{I \uparrow}$ for various Rabi-couplings, $g_{I \downarrow}$ and mass ratios (see legends). A value of $C^{(2)}_{\uparrow \downarrow}<1$ signifies suppression of ferromagnetic medium correlations and thus the generation of the magnetic polaron. 
(c) Comparison between the mean-field and many-body predictions shows negligible deviations and therefore the spin-spatial correlations are important. 
Other parameters are the same as in Fig.~\ref{fig:1bden_immiscible}.}
\label{fig:spin_corel} 
\end{figure*}

\subsection{Dressing cloud and waveform of the magnetic polaron}\label{dressing_cloud}

To reveal the imprint of the impurity into the spatial configuration of the spinor gas we operate in the co-moving impurity frame. 
This is the natural frame of reference to measure the magnonic or phononic dressing cloud since it avoids effects stemming from the dispersion of the impurity within the confining potential. 
Such a procedure has been already successfully implemented in the experiment for monitoring the internal structure of magnetic Fermi polarons~\cite{koepsell2019imaging}. 
Focusing on the bosonic case described above and in order to extract the spatial distribution of the polaron dressing cloud we consider the following two-body density ansatz
\begin{equation}
  \rho^{(2)}_{I a} (x_a,x_I) \approx \left[ \rho^0_a(x_a) + u_a(x_a-x_I)  \right] \rho^0_I(x_I),  
  \label{2bden_ansatz}
\end{equation}
where $\rho^0_a(x)$, $\rho^0_I(x)$ are the background\footnote{Notice that in the case $u_a(x_a-x_I) \neq 0$, $\rho^0_a(x_a) \neq \rho_a(x_a) = \int \mathrm{d}x_I~\rho^{(2)}_{Ia}(x_a,x_I)$ and $\rho^0_I(x_I) \neq \rho_I(x_I) = \int \mathrm{d}x_a~\rho^{(2)}_{Ia}(x_a,x_I)$ hold.} densities of the spin-$a$ component and the  impurity respectively.
These do not directly contribute to the binding of the impurity to the host excitations but they account for the density inhomogeneity originating from the harmonic confinement.
The function $u_a(x_a-x_I)$ captures the impurity dressing by the atoms of its host. 
It is assumed to be solely a function of $x_a-x_I$, due to the short range character of the impurity-host interactions. 
As such, it results in significant variations of the BEC density only for $|x_a - x_I| \sim \xi_a$, where $\xi_a$ is the healing length of the spin-$a$ component.
For a homogeneous system $\rho^0_a(x) \to N_a/L$ and $\rho^0_I(x) \to 1/L$, where $L$ is the length of the system, and thus $\rho^{(2)}_{I a} (x_a,x_I)$ reduces to the form expected within the Lee-Low-Pines mean-field description~\cite{grusdt2017bose,koutentakis2021pattern,jager2021stochastic}.
Therefore, we anticipate that Eq.~(\ref{2bden_ansatz}) describes adequately the structure of $\rho^{(2)}_{I a} (x_a,x_I)$ only in the case that the local density approximation is justified, i.e. $\xi_a \ll R_{\rm out}$.

Provided that Eq.~(\ref{2bden_ansatz}) is valid we can extract information regarding the dressing cloud of the impurity by employing the impurity-spin-$a$ correlation function in the comoving frame $x_r=x_{a}-x_I$, namely 
\begin{equation}
\begin{split}
\bar{\rho}^{(2)}_{I a}(x_r) &\equiv \int \mathrm{d}x_I \rho^{(2)}_{I a}(x_I+x_r,x_I) \\
& \approx u_a(x_r) + \int \mathrm{d}x_I~\rho^0_a(x_I + x_r, x_I) \rho^0_I(x_I).
\end{split}
\label{comoving_correlation}
\end{equation}
Then $\Delta \bar{\rho}^{(2)}_{I a}(x_r)=\bar{\rho}^{(2)}_{I \uparrow}(x_r)-\bar{\rho}^{(2)}_{I \downarrow}(x_r)$ is related to the existence of spin-wave excitation (magnon) dressing, captured by $u_\uparrow(x_r) - u_\downarrow(x_r)$. Moreover, $\bar{n}^{(2)}_{IB}(x_r) = \sum_{a=\uparrow, \downarrow}\bar{\rho}^{(2)}_{I a }(x_r)$ is associated to the presence of phonons. 
Note that for the case of a miscible spinor host (here achieved for $\Omega_R =5 \omega$) it holds that $\langle \hat{S}_z (x_a) \rangle = 0$ for $g_{I \uparrow}=0$.
Therefore, we expect that any deviation in the magnetization of the system for $g_{I \uparrow} \neq 0$ stems from the magnetic dressing cloud of the impurity.
This implies that $\rho^0_\uparrow(x) = \rho^0_\downarrow(x)$ and consequently $\Delta \bar{\rho}^{(2)}_{I a}(x_r) = u_{\uparrow}(x_r) - u_{\downarrow}(x_r)$. 
In this way, $\Delta \bar{\rho}^{(2)}_{I a}(x_r)$ captures the waveform of the standing spin-wave dressing cloud.
We present $\Delta \bar{\rho}^{(2)}_{I a}(x_r)$ for $\Omega_R=5\omega$ with respect to $g_{I \uparrow}$ in Fig.~\ref{fig:dressing} (a). 
It features $\Delta \bar{\rho}^{(2)}_{I a}(x_r) < 0$ ($\Delta \bar{\rho}^{(2)}_{I a}(x_r)>0$) for $g_{I \uparrow} > 0$ ($g_{I \uparrow} > 0$) indicating the respective magnetization tendency. 
The shape of $\Delta \bar{\rho}^{(2)}_{I a}(x_r)$ is found upon fitting to be well-described by 
\begin{equation}
\Delta \bar{\rho}^{(2)}_{I a}(x_r)=-\textrm{sign}(g_{I \uparrow} + g_{I \downarrow})A_{\rm mag} \sech^2\bigg(\frac{x_r}{w}\bigg),\label{wave_repul_magn}
\end{equation}
with $A_{\rm mag}$ ($w$) denoting its amplitude (width). 
We expect that operating in the Lee-Low-Pines framework one should be able to extract Eq.~(\ref{wave_repul_magn}) as the solution of the magnetic polaron dressing cloud, an analysis that will be addressed in a future work. 

\begin{figure*}[ht]
\includegraphics[width=0.9\textwidth]{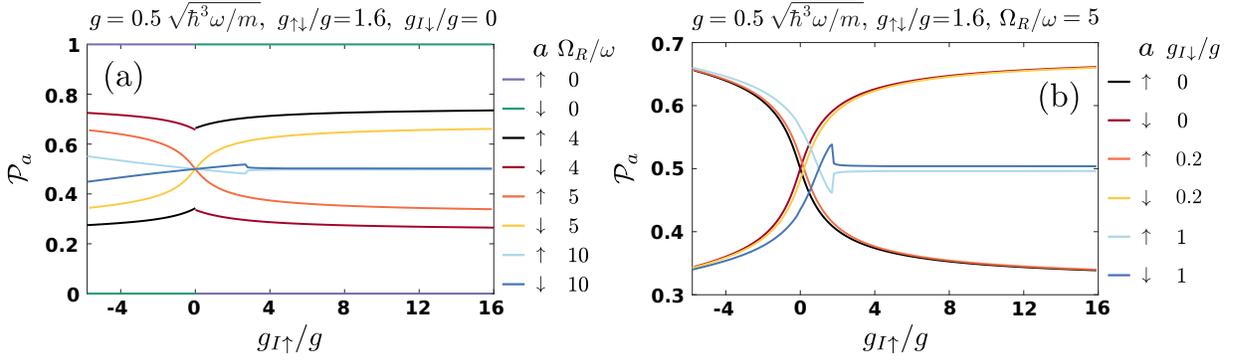}
\caption{Spin transfer between the components of an immiscible medium interacting with an impurity for varying $g_{I \uparrow}$ and distinct Rabi-couplings $\Omega_R$, or interactions $g_{I \uparrow}$, $g_{\uparrow \downarrow}$ (see legends). 
A finite atom migration, i.e. $\Delta \mathcal{P} = \mathcal{P}_{\uparrow} - \mathcal{P}_{\downarrow} \neq 0$, demonstrates the emergence of spin-wave excitations leading to the polaron dressing. }
\label{fig:transfer_immiscible} 
\end{figure*} 

On the other hand, it is more involved to extract the phononic dressing cloud as the non-vanishing contribution from the last term of Eq.~(\ref{comoving_correlation}) implies
$u_{\uparrow}(x_r) + u_{\downarrow}(x_r) \neq \bar{n}_{IB}^{(2)}(x_r)$.
To account for this correction we employ the following spin-independent pair-correlation ansatz 
\begin{equation}
\begin{split}
    \bar{n}^{(2)}_{IB}(x_r) = n_{\rm TF} (x_r) - \underbrace{A_{\rm ph} n_{\rm TF} (x_r) \sech^2\bigg(\frac{x_r}{w'}\bigg)}_{= u_\uparrow(x_r) + u_\downarrow(x_r) },
\end{split}
\label{wave_attr_magn}
\end{equation}
Here, $w'$ refers to the width of the polaron and $A_{\rm ph}$ is the amplitude of the phonon dressing cloud. In the above expression we have assumed a 
TF (Gaussian) background density for the BEC (impurity). 
With this choice the background two-body density, $n_{\rm TF}(x_r)$, reads
\begin{equation}
\begin{split}
    &n_{\rm TF}(x_r) = \int {\rm d}x_I~\left[\sum_{\alpha \in \{ \uparrow, \downarrow \}} \rho^{\rm TF}_{\alpha}(x_I+x_r) \right] \rho^{l}_I(x_I)\\
    &= \frac{m \omega^2}{4 g_{\rm eff}} \bigg[ \left( {\rm erf} \frac{x_r + R_{\rm out}}{l} + {\rm erf} \frac{x_r - R_{\rm out}}{l} \right)\\
    &\times \left(R_{\rm out}^2 -\frac{l^2}{2}  -x_r^2 \right) + \frac{l (R_{\rm out} - x_r)}{\sqrt{\pi}} e^{- \frac{(R_{\rm out} + x_r)^2}{l^2}} \\
    &+ \frac{l (R_{\rm out} + x_r)}{\sqrt{\pi}} e^{- \frac{(R_{\rm out} - x_r)^2}{l^2}} \bigg],
    \end{split}
\end{equation}
where $\rho^{\rm TF}_{\alpha}(x)$ corresponds to the TF profile of the spin-$\alpha$ BEC component (see also Appendix~\ref{app:mean_field_binary}) and $\rho^{l}_I(x_I) = \frac{1}{l \sqrt{\pi}} e^{-\frac{x^2}{2 l^2}}$.
Also, the fitting parameters $R_{\rm out}$ and $g_{\rm eff}$ account for the width and height of the BEC density profile, while $l$ corresponds to the width of the impurity density. Recall that in the un-trapped case using the Lee-Low-Pines transformation leads to a similar form to Eq.~(\ref{wave_attr_magn}) for the dressing cloud of the Bose polaron~\cite{koutentakis2021pattern,jager2021stochastic,panochko2019mean}. 

Upon fitting the ansatz of Eq.~(\ref{wave_repul_magn}) to $\Delta \bar{\rho}^{(2)}_{I a}(x_r)$ and the one of  Eq.~(\ref{wave_attr_magn}) into $\sum_{a=\uparrow, \downarrow}\bar{\rho}^{(2)}_{I a }(x_r)$ we determine the width and amplitude of the magnetic polaron and the polaron respectively. 
Next, in order to discern the dressing cloud stemming from the magnons and the phonons we find the number of medium atoms lying in the waveforms $u_{\uparrow}(x_r) \pm u_{\downarrow}(x_r)$, see Eq.~(\ref{wave_repul_magn}) and Eq.~(\ref{wave_attr_magn}). 
These populations, $N_{{\rm mag}}=\int dx_r \Delta \bar{\rho}_{I a}^{(2)}(x_r)$ and $N_{{\rm ph}}=\int dx_r (\bar{n}^{(2)}_{IB}(x_r)-n_{\rm TF} (x_r))$ depicted in Figs.~\ref{fig:dressing} (b), (c) feature an increasing tendency for a larger magnitude of impurity-spin-$\uparrow$ interactions testifying quasiparticle formation. 
It can also be inferred that independently of $g_{I \uparrow}$ the magnonic excitations prevail over the phononic ones. 
This means that the emergent quasiparticle, being genuinely dressed by an admixture of magnons and phonons, possesses a dominant magnetic component. 
Particularly, the magnetic dressing cloud acquires a maximal value of about $30\%$ of the medium atoms around $\Omega_R^{\rm crit} \approx 4.8$, whilst the phonon branch is at most $3.5\%$.

For $\Omega_R=10 \omega >\Omega_R^{\rm crit}$ the magnonic excitation branch is again pronounced as compared to the phononic one especially for $g_{I \uparrow}<0$, while both branches are suppressed for $g_{I \uparrow}/g>3.8$ where phase-separation occurs. 
Naturally, if $\Omega_R=0$ magnetic dressing is diminished for every $g_{I \uparrow}$, while the phononic one is finite for $g_{I \uparrow}<0$ since only in this case the impurity lies within the spin-$\uparrow$ host. 
Regarding $\Omega_R=40 \omega \gg \Omega_R^{\rm crit}$, we observe that $N_{\rm mag}$ is suppressed. 
In this scenario the spin degrees-of-freedom are almost frozen because of the large energy gap for exciting spin-waves~\cite{abad2013study}.   
Concluding, from the above we can deduce that the magnonic cloud should be easier experimentally detectable when compared to its phononic counterpart.

\subsection{Broken spin-order and spin-spin correlations}\label{spin_corel} 

To inspect the rise of spin-fluctuations in the medium and thus attest the emergence of the magnetic polaron we track the spin-spin correlation~\cite{koutentakis2019probing} 
\begin{equation}
C^{(2)}_{\uparrow \downarrow}=\frac{\braket{ \Psi | \hat{S}^2 | \Psi}-3\hbar^2N}{\hbar^2 N(N-1)}. \label{spin_corel_func}
\end{equation}
Here, $\hat{S}^2$ is the total-spin operator of the system defined in Eq.~(\ref{total_spin}). 
This correlation function probes the alignment among two spins and dinstinguishes ferromagnetic $C^{(2)}_{\uparrow \downarrow} \approx 1$, antiferromagnetic $C^{(2)}_{\uparrow \downarrow} \approx -1$ and paramagnetic $C^{(2)}_{\uparrow \downarrow}= 0$ spin configurations\footnote{A fully  ferromagnetic gas has  $\braket{\hat{S}^2}=N(N+2)$ for every $N$. The minimal value of   $\braket{\hat{S}^2}=0$ does not allow for a perfectly anti-ferromagnetic configuration, since $C^{(2)}_{\uparrow \downarrow}=-3/(N-1)$. Indeed, it is impossible that every pair of spins is anti-oriented which suppresses the perfect anti-ferromagnetic order.}. 
It is showcased in Fig.~\ref{fig:spin_corel} (a) for immiscible spin-spin interactions and several Rabi-couplings as a function of the impurity-spin-$\uparrow$ interaction strength. 
As it can be readily seen, for suppressed Rabi-coupling ($\Omega_R=0$) the medium remains ferromagnetic (i.e. $C^{(2)}_{\uparrow \downarrow}=1$) irrespectively of $g_{I \uparrow}$. 
Recall that the interacting eigenstate of the multicomponent setting corresponds to the one where all atoms are in the spin-$\uparrow$ ($\downarrow$) state for $g_{I \uparrow}>0$ ($g_{I \uparrow}<0$). 
In this scenario, a polaron dressed by the phononic excitations of the bosonic bath occurs as long as $g_{I \uparrow}<g+g_{\uparrow \downarrow}$. 
Otherwise, an impurity-medium phase-separation takes place evincing the polaron decay~\cite{mistakidis2020many}. 

In sharp contrast, switching on the Rabi-coupling results generally to the suppression of the ferromagnetic order as evidenced by the reduction of $C^{(2)}_{\uparrow \downarrow}$ for increasing $\abs{g_{I \uparrow}}$ [Fig.~\ref{fig:spin_corel} (a)]. 
Particularly, spin correlations become enhanced within the critical magnetization region $\Omega_R^{\rm crit} \approx n_{B}(0)(g_{\uparrow \downarrow}-g_{\uparrow \uparrow})\approx 4.8$ where spin-mixing is dominant [Appendix~\ref{app:mean_field_binary}]. 
This behavior of $C^{(2)}_{\uparrow \downarrow}$ testifies the existence of the magnetic polaron being dressed by the spin-wave excitations of the spinorial medium~\cite{ashida2018many}, see also the discussion in Section~\ref{spin_transf_immiscible}. 
Notably, $C^{(2)}_{\uparrow \downarrow} \approx 1$ for $\Omega_R \gg \Omega_{R}^{cr}$, see e.g. $\Omega_R/\omega=10$, where impurity spin-$\uparrow$ spatial separation occurs.  
On the other hand, $C^{(2)}_{\uparrow \downarrow} \neq 1$ for $g_{I \uparrow}<0$ ($g_{I \uparrow}>0$) supporting the occurrence of a self-bound attractive (stable repulsive) magnetic Bose polaron. 

A similar to the above-described response of the spin-fluctuations takes place for a heavy impurity ($m_I \gg m_B$) or finite impurity spin-$\downarrow$ couplings, see Fig.~\ref{fig:spin_corel} (b). 
Notice that $C^{(2)}_{\uparrow \downarrow} \to 1$ for $g_{I \downarrow}/g>1.6$ since then a phase-separation between the impurity and the spinor medium is favored as long as $g_{I \uparrow}> g+g_{\uparrow \downarrow}-g_{I \downarrow}$.

\begin{figure*}[ht]
\includegraphics[width=0.8\textwidth]{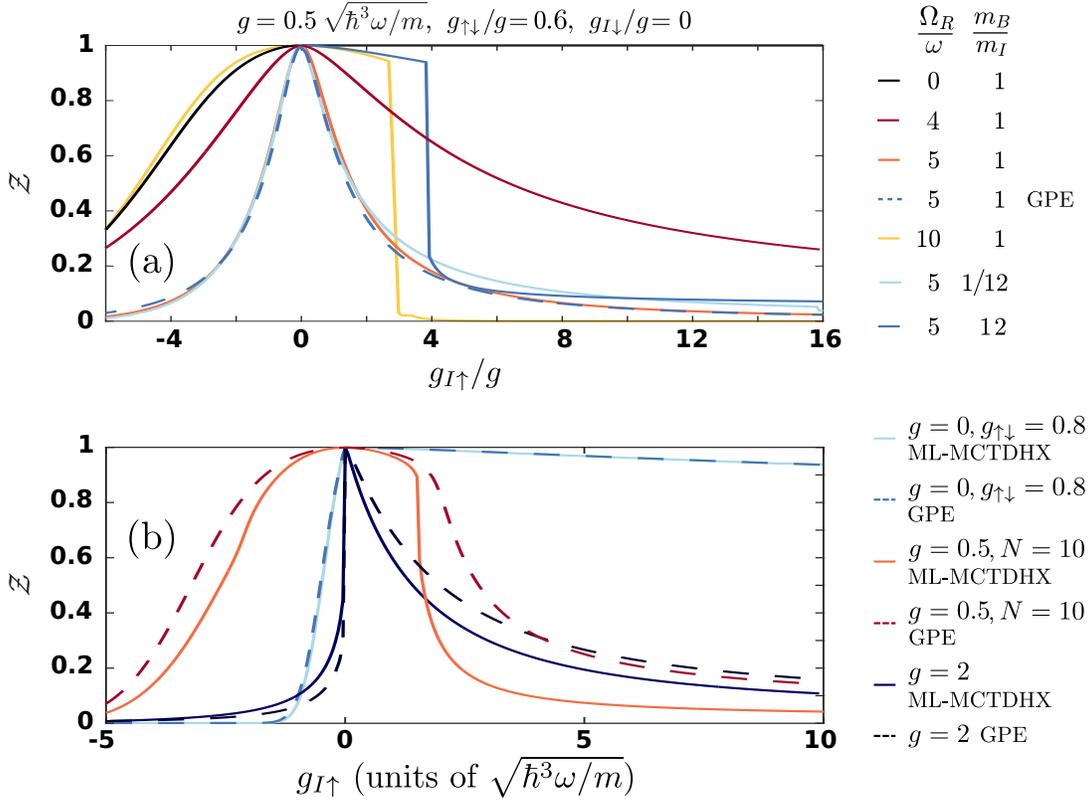}
\caption{(a) Magnetic polaron residue in the case of an immiscible interacting medium with varying $g_{I \uparrow}$ and different $\Omega_R$ (see legend). 
A decrease of $\mathcal{Z}$ for finite $\abs{g_{I \uparrow}}$ shows the formation of an attractive ($g_{I \uparrow}<0$) or a repulsive ($g_{I \uparrow}>0$) magnetic polaron state for $\Omega_R \neq 0$. The fact that $\mathcal{Z} \to 0$ for increasing $g_{I \uparrow}>g$ hints towards the catastrophe of the magnetic polaron. The predictions of the mean-field theory are in good agreement with the many-body residue outcome for weak intraspecies bath interactions.  
(b) The mean-field framework overestimates the magnetic polaron residue for a strongly interacting medium, compare $\mathcal{Z}$ as obtained within the variational approach and the GPE. 
The remaining system parameters are shown in the legends. }
\label{fig:residue} 
\end{figure*}

\begin{figure}[ht]
\includegraphics[width=0.9\columnwidth]{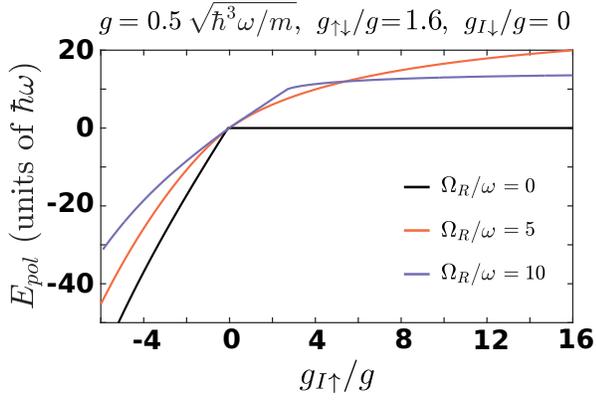}
\caption{Energy of the magnetic polaron for an immiscible interacting medium as a function of $g_{I \uparrow}$ for different Rabi-couplings (see legend). 
Attractive ($g_{I \uparrow}<0$) and repulsive ($g_{I \uparrow}>0$) magnetic polaron branches exist for non-zero $\Omega_R$ characterized by a negative and a positive energy respectively. 
Stronger binding occurs in the vicinity of the miscibility threshold, $\Omega_R\approx 4.8$. 
Other system parameters are provided in the legends. }
\label{fig:energies} 
\end{figure}

\subsection{Spin transfer processes}\label{spin_transf_immiscible}

Having exemplified the role of spin-fluctuations caused exclusively by the impurity-medium interactions we next aim to unravel the accompanied spin-demixing mechanisms. 
These refer to the migration of spin-$\uparrow$ to spin-$\downarrow$ particles and vice versa leading ultimately to interaction dependent spin configurations. 
The latter naturally provide further evidence of the presence of spin-wave excitations identified in Fig.~\ref{fig:dressing} (a). 
These are quantified herein by the portion of the spin-flipped atoms  with respect to the $g_{I \uparrow}=0$ configuration. To capture the spin-demixing of the bosonic medium, due to $g_{I \uparrow} \neq 0$, for various Rabi-couplings $\Omega_R$ we monitor the fraction of bosons in each spin-$a$ component 
\begin{equation}
\mathcal{P}_{a}(\Omega_R;g_{aa'},g_{Ia})=\frac{\langle N_{a} (\Omega_R;g_{aa'},g_{Ia})\rangle}{N},\label{spin_frac} 
\end{equation}
where $N=N_{\uparrow}+N_{\downarrow}$. 
The respective fraction of spin-$a$ atoms for several $\Omega_R$ is depicted in Fig.~\ref{fig:transfer_immiscible}(a). In the miscible regime, i.e. $\Omega_R >\Omega_R^{{\rm crit}} \approx 4.8 \omega$, it holds that $\mathcal{P}_{a}(\Omega_R;g,g_{Ia}=0)=1/2$, i.e. $\Delta \mathcal{P}=\mathcal{P}_\uparrow - \mathcal{P}_\downarrow = 0$.
For $g_{I\uparrow} >0$ ($g_{I\uparrow} <0$) $\Delta P$ decreases (increases).
In the case of $\Omega_R \gg \Omega_R^{{\rm crit}}$ and $g_{I \uparrow} > g+ g_{\uparrow \downarrow}$, $\Delta \mathcal{P}=0$ due to phase-separation. 
In the immiscible scenario, $\Omega_R < \Omega_R^{\rm crit}$, $\Delta \mathcal{P} \neq 0$ even for $g_{I \uparrow} =0$ and it follows the same behavior as in the miscible regime with $g_{I \uparrow}$.
For $\Omega_R=0$, $\Delta \mathcal{P}=\pm 1$ because the spinor two-component host reduces to a single-component since $\Omega_R < \Omega_R^{{\rm single}}$ [see also Fig.~\ref{fig:schematic_phase}]. 

This response of $\Delta \mathcal{P}$ in the miscible regime (here $\Omega_R =5 \omega$) does not alter also for a finite value of $g_{I \downarrow}$ as long as $g_{I \downarrow}<g$, see Fig.~\ref{fig:transfer_immiscible} (b). 
Of course, due to $g_{I \downarrow}\neq 0$ the population balance scenario is achieved for $g_{I \uparrow}=g_{I \downarrow}$. 
This becomes more pronounced for $g_{I \downarrow}/g=1$ where we also observe that $\Delta \mathcal{P} \to 0$ when $g_{I \uparrow}>g_{\uparrow \downarrow}$, see Appendix~\ref{app:effective_pot}. 
The fact that $\Delta \mathcal{P} \to 0$ is essentially a manifestation of the impurity-medium phase-separation for these non-negligibly repulsive values of $g_{I \downarrow}$. 
Consequently, the impurity lies outside of the BEC and thus spin-demixing is diminished. 
Such a suppression of $\Delta \mathcal{P}$ for $g_{I \uparrow} + g_{I \downarrow} > g+ g_{\uparrow \downarrow}$ takes equally place upon considering a $\Omega_R \gg \Omega_R^{\rm crit}$ which in general produces a reduction of $\Delta \mathcal{P}$ for each $g_{I \uparrow}$ (not shown).

\subsection{Magnetic polaron residue and energy}\label{res_en} 

The quasiparticle residue~\cite{massignan2014polarons} refers to the overlap between the non-interacting ($g_{Ia}=0$) and the interacting (polaronic) state  
\begin{equation}
\mathcal{Z}=\abs{\braket{\Psi (g_{Ia}) | \Psi (g_{I a}=0)}},\label{residue}
\end{equation} 
with $a=\uparrow,\downarrow$. 
It can be experimentally tracked e.g. via radiofrequency spectroscopy~\cite{cetina2016ultrafast,kohstall2012metastability}. 
The behavior of $\mathcal{Z}$ with respect to $g_{I \uparrow}$ and distinct $\Omega_R$ is presented in Fig.~\ref{fig:residue} (a). 
Apparently, for $\Omega_R=0$ the residue $\mathcal{Z}= 1$ for $g_{I \uparrow}>0$ implying that the impurity is not dressed by its environment, thus corresponding to a free particle. This is in contrast to $g_{I \uparrow}<0$ where $\mathcal{Z}$ reduces for stronger attractions being a consequence of the attractive polaron dressed by the phononic excitations of the bosonic medium.  
Turning to larger $\Omega_R$, in the vicinity of $\Omega_R^{\rm crit}$, we find that $\mathcal{Z}$ reduces for increasing $\abs{g_{B \uparrow}}$ signifying a tendency towards a completely deformed interacting state. 
For repulsive $g_{I \uparrow}$ it is caused by the phase-separation, while for attractive interactions it stems from the impurity-medium bound state. 
This overall behavior of $\mathcal{Z}$ appears to be similar for a heavy impurity. 
For $\Omega_R>\Omega_R^{\rm crit}$ the polaronic residue is dramatically different. 
Focusing on $g_{I \uparrow}<0$ it decreases for larger attractions but it is always larger than for $\Omega_R/\omega=5$. 
However, repulsive impurity-spin-$\uparrow$ interactions result in a smooth decrease of $\mathcal{Z}$ for $g_{I \uparrow}<g+g_{\uparrow \downarrow}$. 
Otherwise a sharp reduction of $\mathcal{Z}\to 0$ takes place which is associated with the phase-separation among the impurity and the spin-$\uparrow$ state. 
A similar response is observed for a heavy medium and $\Omega_R/\omega=5$. 

We then compare our variational findings with the results from the mean-field approximation. 
Such a direct comparison is especially motivated by recent studies on the 1D Bose polaron which have been argued that its ground state characteristics can be described (at least to some extent~\cite{brauneis2022artificial,brauneis2021impurities,mistakidis2022cold}) within a mean-field framework~\cite{panochko2019mean}. 
This issue has been carefully benchmarked, for instance, against Quantum Monte carlo techniques~\cite{will2021polaron}, the Lee-Low-Pines transformation~\cite{jager2020strong,koutentakis2021pattern} as well as the ML-MCTDHX and the flow equation (IM-SRG) methods~\cite{brauneis2022artificial,brauneis2021impurities}. 
The respective mean-field approximation for our system is described by the following system of three coupled Gross-Pitaevskii equations   
\begin{equation}
    \begin{split}
    &\bigg[ - \frac{\hbar^2}{2 m_B} \frac{\mathrm{d}^2}{\mathrm{d} x^2} + \frac{1}{2} m_B \omega_B^2 x^2  -\mu_B + \tilde{g}_{\uparrow \uparrow} \left| \psi_{\uparrow}(x) \right|^2 \\
    &+ \tilde{g}_{\uparrow \downarrow} \left| \psi_{\downarrow}(x) \right|^2 +  g_{I \uparrow} \left| \psi_{I}(x) \right|^2 \bigg] \psi_{\uparrow}(x) + \frac{\Omega_R}{2} \psi_{\downarrow}(x) =0 \\
    &\bigg[ - \frac{\hbar^2}{2 m_B} \frac{\mathrm{d}^2}{\mathrm{d} x^2} + \frac{1}{2} m_B \omega_B^2 x^2  -\mu_B + \tilde{g}_{\downarrow \downarrow} \left| \psi_{\downarrow}(x) \right|^2 \\
    &+ \tilde{g}_{\uparrow \downarrow} \left| \psi_{\uparrow}(x) \right|^2 +  g_{I \downarrow} \left| \psi_{I}(x) \right|^2 \bigg] \psi_{\downarrow}(x) + \frac{\Omega_R}{2} \psi_{\uparrow}(x) =0 \\
    &\bigg[ - \frac{\hbar^2}{2 m_I} \frac{\mathrm{d}^2}{\mathrm{d} x^2} + \frac{1}{2} m_I \omega_I^2 x^2  -\mu_I + g_{I \uparrow} \left| \psi_{\uparrow}(x) \right|^2 \\
    &+ g_{I \downarrow} \left| \psi_{\downarrow}(x) \right|^2 \bigg] \psi_{I}(x) =0,
    \end{split}
    \label{GrossPitaevskii_3C}
\end{equation}
where $\tilde{g}_{a a'}=(1-1/N) g_{a a'}$. 
The chemical potential of the medium $\mu_B$ is inherently related with the particle number\footnote{$\mu_B$, $\mu_I$ correspond to the solution of 
two algebraic equations, namely $\int \mathrm{d}x~| \psi_\uparrow(x;\mu_B,\mu_I) |^2 + | \psi_\downarrow(x;\mu_B,\mu_I) |^2 = N$ and $\int \mathrm{d}x~| \psi_I(x;\mu_B,\mu_I) |^2 = 1$,
where $\psi_\sigma(x;\mu_B,\mu_I)$ is the wave function obtained from Eq.~(\ref{GrossPitaevskii_3C}) as a function of $\mu_B$, $\mu_I$.} $N=N_{\uparrow}+N_{\downarrow}$, see also Appendix~\ref{app:effective_pot}. 
Importantly, the mean-field framework accounts for the hybridization of the spin and spatial degrees-of-freedom ignoring the excited state contributions or effects originating from quantum fluctuations~\cite{mistakidis2022cold}. 
Comparing the quasiparticle weight between the many-body approach [Appendix~\ref{app:variational_treat}] and the mean-field theory [Eq.~(\ref{GrossPitaevskii_3C})] reveals only small deviations for weak boson-boson interactions, see Fig.~\ref{fig:residue} (a). 
From this we can conclude that spin-spatial correlations play the dominant role in the behavior of $\mathcal{Z}$ and thus on the generation of the magnetic polaron. 
Spatial correlations giving rise to deviations from the mean-field picture become relevant for stronger boson-boson interactions. 

Indeed, deviations from the mean-field become in particular noticeable for repulsive impurity-medium interactions and around $\Omega_R^{\rm crit}$, implying that in this region spatial correlations become non-negligible. 
The imprint of spatial correlations on $\mathcal{Z}$ is especially pronounced by considering stronger interparticle interactions, compare the many-body and the mean-field results in Fig.~\ref{fig:residue} (b). 
In particular, the presence of correlations lead to smaller residue for repulsive $g_{I \uparrow}$ due to the superposition of the many-body wave function as compared to the mean-field one. 
For attractive $g_{I \downarrow}$ the situation is reversed because the impurity is less spatially localized in the mean-field case.

Since the mean-field solution is adequate for weak medium interactions, we have also studied the behavior of $\mathcal{Z}$ for substantially larger atom numbers. 
For this investigation we systematically approach the thermodynamic limit\footnote{
Namely, we keep constant the healing length of the medium, i.e. $\xi \propto 1/\sqrt{\mu_B}$ [Eq.~(\ref{eq_chem_pot})], while simultaneously increasing the medium density, $n_B(0) \propto \mu_B/(g_{\uparrow \downarrow}+g)$ which is achieved by fixing $g_{aa'} N={\rm const}$ while increasing $N$ and $N_I$ by the same factor.} observing that $\mathcal{Z} \sim | \braket{\psi_\uparrow| \psi_{\uparrow 0}} + \braket{\psi_\downarrow| \psi_{\downarrow 0}} |^{N_B} | \braket{\phi_I | \phi_{I0}} |^{N_I}$, where $| \phi_{\sigma} \rangle$ and $| \phi_{\sigma 0} \rangle$, with $\sigma \in \{ \uparrow, \downarrow, I \}$ are the mean-field order parameters for $g_{I \uparrow} \neq 0$ and $g_{I \uparrow} = 0$ respectively.
Thus, $\mathcal{Z}(N_B,N_I) \approx \tilde{\mathcal{Z}}^{\frac{N_B}{100}} \to 0$ for large $N_B$ and $N_I$, where $\tilde{\mathcal{Z}}$ corresponds to the residue for $N=100$. This exponentially decaying behavior of $\mathcal{Z}$ manifests the Anderson catastrophe of the magnetic polaron in the thermodynamic limit in 1D~\cite{anderson1967infrared,massignan2014polarons}. 
It further supports the generalization of our results and their experimental detection in the large atom number limit as long as the impurities density is low to ensure that their interactions are negligible.  

The energy of the quasiparticle is naturally determined by the energy difference among the interacting impurity-medium setting and the non-interacting one    
\begin{equation}
E_{{\rm{pol}}}=\braket{\hat{H}(g_{Ia})}-\braket{\hat{H}(g_{Ia}=0)}.\label{polaron_energy}
\end{equation}
As it can be verified by inspecting Fig.~\ref{fig:energies} the polaron energy shows a continuously increasing (decreasing) trend for stronger repulsive (attractive) $g_{I \uparrow}$ as long as $\Omega_R\neq 0$. Moreover, for $\Omega_R>\Omega_R^{\rm crit}$ and in the case of $g_{I \uparrow}>g+g_{\uparrow \downarrow}$ the energy saturates due to the impurity-spin-$\uparrow$ phase-separation process [see also Fig.~\ref{fig:1bden_immiscible}]. 
The saturation value of the polaron energy caused by the impurity-medium phase-separation can also be predicted within $V_{\rm eff}(x)$ [see Eq.~(\ref{eq:effective_pot_immiscible_TF}) in Appendix~\ref{app:effective_pot}] and it corresponds roughly to $E_{\rm ph. s.} \sim (1/2) m_I \omega_I^2 R_{\rm out}^2$ with $R_{\rm out}$ being the TF radius of the interacting component. 
However, for $\Omega_R \approx \Omega_R^{\rm crit}$ the magnetic polaron energy does not exhibit an upper bound due to the absence of phase-separation.
Notice that in the case of $\Omega_R=0$ the energy $E_{{\rm{pol}}}=0$ in the repulsive $g_{I \uparrow}>0$ regime since there is no dressing, whilst $E_{{\rm{pol}}}<0$ for $g_{I \uparrow}<0$ due to the self-bound attractive Bose polaron.

\section{Conclusions and outlook}\label{conclusions} 

We have investigated the ground state properties of a structureless impurity embedded in a spin-$1/2$ Bose gas. 
The interaction of the impurity with one spin component is switched on, leading to spin-wave excitations of the host atoms and formation of magnetic Bose polarons. 
To evaluate beyond mean-field correlations in the magnetic polaron formation, we compare results obtained within the variational approach and the three-component Gross-Pitaevskii equations. 
The spin-spatial correlations play the dominant role for weak boson-boson interactions, while spatial correlations become important for stronger interparticle couplings, invalidating the mean-field treatment.
An effective potential for the impurity immersed in the bosonic bath is constructed to elucidate the polaron characteristics. 
Interestingly in the magnetic polaron regime the impurity is confined within an effective attractive potential well originating from the host excitations it triggers. 
Particularly, regarding the repulsive branch this emergent attraction stabilizes the quasi-particle against impurity-bath phase separation. 

The phase diagram of the 1D magnetic polaron, as a function of impurity-medium, spin-spin interactions, and Rabi coupling among the spin components is calculated.
The new phases include attractive and self-bound repulsive magnetic polaron configurations, and impurity-medium phase-separated regions where the quasiparticle decays. 
The transition from non-magnetic to magnetic polaron states strongly depends on the Rabi coupling. 
The spin-wave excitations emerge as a localized disturbance of the host magnetization, affecting the occupation of the individual spin components. This mechanism is tunable by varying the impurity-medium interactions while keeping all other system parameters fixed. 
In the absence of impurity, it is explicated that the binary magnetic gas can be efficiently driven through a miscible to spin-segregated phase via tuning its Rabi-coupling. 
It is in the vicinity of this phase transition where the magnetic properties of the polaron become more prominent. 

We show that spin-spin correlations are affected with finite impurity-medium coupling, offering the opportunity to use impurity-spin interaction to control the local bath spin order. 
To gain further insights into the magnetic polaron states, we calculate the residue, showing a decreasing trend for large impurity-medium coupling. 
The presence of spatial correlations for strong boson-boson interactions results in a suppressed residue behavior. 
Moreover, by inspecting the polaron energy it is shown to be negative for attractive impurity-medium couplings indicating the formation of an impurity-medium bound state. 
The latter is less bound for larger Rabi coupling, while for repulsive interactions it experiences an increasing tendency. 

It would be intriguing to develop an analytical understanding of the impact of correlations leading to the renormalization of the effective potential experienced by the impurity within the magnetic polaron regime.
Another immediate prospect is to examine the quench dynamics of the magnetic polaron following a time-dependent ramp of the intensity of the radiofrequency field (Rabi coupling) in order to examine the possibility of spin domain formation. Along these lines, it is important to emulate radiofrequency or Ramsey spectroscopy in the spinor medium for unveiling the many-body properties of its collective excitations, such as spin-waves.
Another interesting direction would be to study the induction of spin-order in the presence of a spin-orbit coupled impurity atom. 
The generalization of our findings to two-dimensions where long-range effective spin-order, exhibiting also anisotropic character, is a worthy further pursuit.

\section*{Acknowledgements} 
S. I. M. and H.R.S. acknowledge support from the NSF through a grant for ITAMP at Harvard University. 
G.M.K. and P. S. have been funded by 
the Cluster of Excellence “Advanced Imaging of Matter” of 
the Deutsche Forschungsgemeinschaft (DFG)-EXC 2056-project ID 390715994. 
F. G. acknowledges funding by the Deutsche Forschungsgemeinschaft (DFG, German Research Foundation) under 
Germany’s Excellence Strategy -- EXC-2111 -- 390814868. 

\appendix

\section{Impurity in a miscible spinor Bose medium}\label{app:miscible_bath}

\begin{figure*}[ht]
\includegraphics[width=1.0\textwidth]{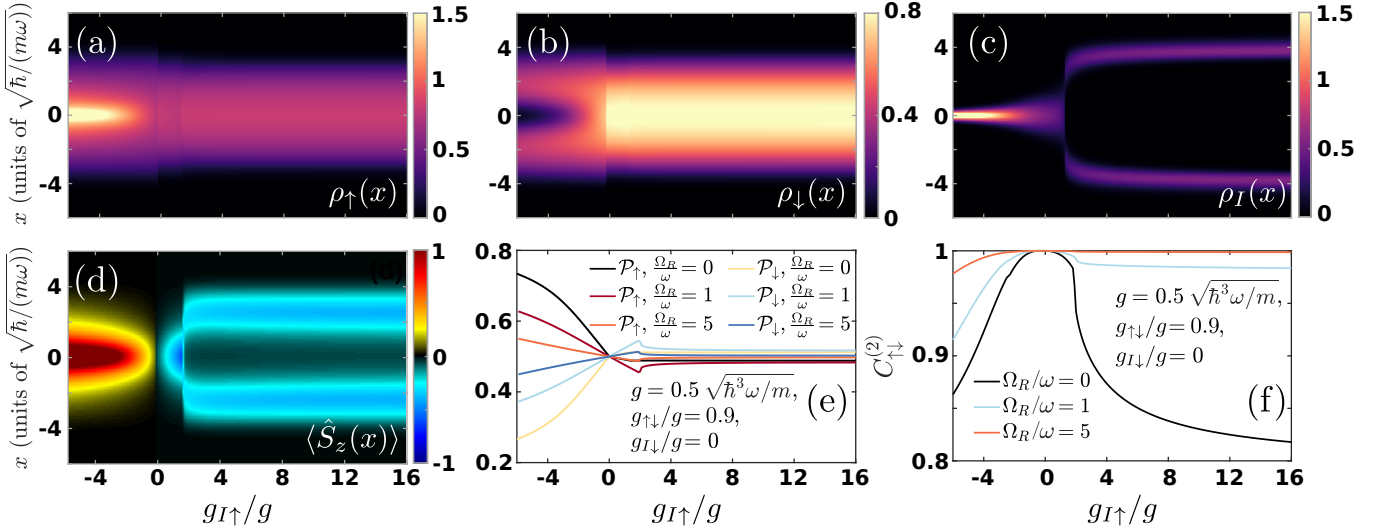}
\caption{Ground state of an impurity immersed in a miscible interacting spinor Bose gas characterized by Rabi-coupling $\Omega_R=0$. The densities of (a) the spin-$\uparrow$ and (b) the spin-$\downarrow$ components as well as (c) the impurity are shown for varying impurity-spin-$\uparrow$ coupling. 
(d) The respective spatially resolved spin-fluctuations. 
(e) Spin populations of the individual components of a miscible medium interacting with an impurity upon tuning $g_{I \uparrow}$ for different $\Omega_R$ (see legends). 
(f) the spn-spin correlation function. 
The multicomponent interacting Bose system with $g_{\uparrow \uparrow}=g_{\downarrow \downarrow}\equiv g =0.5$, $g_{\uparrow \downarrow}/g=0.9$ and $g_{I \downarrow}=0$ is harmonically trapped with $\omega=1$.}
\label{fig:1bden_miscible} 
\end{figure*}

Below, we showcase that the polaron exists also in the case of a miscible interacting spinor medium, however, its magnetic character depends crucially on $\Omega_R$. 
Here, we assume $g=0.5$ and $g_{\uparrow \downarrow}/g=0.9$ while $g_{I \downarrow}=0$. The $\sigma$-component densities are provided in Figs.~\ref{fig:1bden_miscible} (a)-(c) for different impurity-spin-$\uparrow$ interactions and in the case of $\Omega_R=0$. 
For repulsive $g_{I \uparrow}$ we observe that in the region $g_{I \uparrow}< g_{\uparrow \downarrow}+g$ the impurity cloud widens for larger $g_{I \uparrow}$ while remaining within $\rho_{\uparrow}$. Entering $g_{I \uparrow}> g_{\uparrow \downarrow}+g$ the impurity moves to the edges of the TF cloud of the medium, $x = \pm R_{\rm out}$, splitting into two symmetrically placed density branches minimizing its overlap with $\rho_{\uparrow}(x)$. 
However, $\rho_{\downarrow}(x)$ slightly widens as a result of its effective attraction ($g_{I \downarrow}^{\rm eff}$) mediated by the repulsive $g_{I \uparrow}$, see Sec. \ref{distributions_immiscible}. 
The local magnetization in the vicinity of the impurity is negative in this repulsive interaction region manifesting the existence of a magnetic polaron [Fig.~\ref{fig:1bden_miscible}(d)]. 
The magnetic quasiparticle nature is also supported by the occurrence of spin-fluctuations identified since $C_{\uparrow \downarrow}^{(2)} \neq 1$ as shown in Fig.~\ref{fig:1bden_miscible}(f). 
Notably, $C_{\uparrow \downarrow}^{(2)} \to 1$ irrespectively of $g_{I \uparrow}$ for increasing $\Omega_R$ implying that the magnetic character is diminished.
This is expected as for increasing $\Omega_R$ the host becomes progressively more strongly polarized along the spin-$x$ axis [$\ket{P_x} = \bigotimes_{i=1}^N (\ket{\uparrow}_i - \ket{\downarrow}_i)/\sqrt{2}$] and therefore the impurity can hardly induce spin-demixing. 
Consequently, the atoms are distributed in an almost equal fashion between the spin components, see Fig.~\ref{fig:1bden_miscible}(e). 
The latter feature a small population imbalance being more pronounced before the overlap among the impurity and the spin-$\uparrow$ bosons becomes minimal and it is further suppressed for larger $\Omega_R$.   

Turning to attractive $g_{I \uparrow}$, a non-negligible portion of the spin-$\uparrow$ atoms accumulate close to the impurity and a self-bound attractive magnetic polaron forms. 
Simultaneously, $\rho_{\downarrow}(x)$ tends to separate with $\rho_{\uparrow}(x)$ as a result of the repulsive induced interactions ($g_{I \downarrow}^{\rm eff}$) mediated by the attractive $g_{I \uparrow}$. 
Here, $\braket{\hat{S}_z(x)}$ is positive in the location of the impurity [Fig.~\ref{fig:1bden_miscible}(d)] verifying the magnetic character of the polaron which can also be inferred by the fact that $C^{(2)}_{\uparrow \downarrow} \neq 1$ [Fig.~\ref{fig:1bden_miscible}(f)]. 
Once again the reduction of $C_{\uparrow \downarrow}^{(2)}$ is smaller for a larger $\Omega_R$ indicating the suppression of the magnetic polaron character. 
The population transfer is enhanced compared to the repulsive case [Fig.~\ref{fig:1bden_miscible}(e)]. 
This implies that the polarization is not adequate for characterizing spin-spin correlations and vice versa~\cite{koutentakis2019probing}.

\section{Phase structure of the two-component magnetic Bose gas}\label{app:mean_field_binary}

Let us briefly outline the mean-field ground state characteristics of a binary magnetic bosonic gas, analyzed to some extent in Ref.~\cite{abad2013study}. 
A simplified description of the pseudospinor BEC is obtained by relying on the TF approximation in the three-component GPE description [Eq.~(\ref{GrossPitaevskii_3C})]. 
Particularly, the kinetic energy of the medium is neglected leading to the following coupled set of equations for the medium 
\begin{equation}
  \begin{split}
    \bigg(\mu_B-\varepsilon_0(x)-&\frac{1}{2}m_B\omega^2 x^{2} \bigg) \left(
\begin{array}{c}
\psi_{\uparrow}(x)\\
\psi_{\downarrow}(x)
\end{array} \right) \\
&= \underbrace{\frac{1}{2} \left(
\begin{array}{c c}
 \Delta(x) & \Omega_R \\
\Omega_R & \Delta(x)
\end{array} \right)}_{\equiv \hat{H}_s (x)} 
\underbrace{\left(
\begin{array}{c}
\psi_{\uparrow}(x)\\
\psi_{\downarrow}(x)
\end{array} \right)}_{\equiv \langle x | \Psi_B \rangle},
  \end{split}
\label{General_self_consistent}
\end{equation}
where $\varepsilon_0(x) \equiv (\varepsilon_{\uparrow}(x)+\varepsilon_{\downarrow}(x))/2$, $\Delta(x) \equiv \varepsilon_{\uparrow}(x)-\varepsilon_{\downarrow}(x)$ and  $\varepsilon_{\alpha}(x) = \sum_{\beta = \{ \uparrow, \downarrow \}} g_{\beta \alpha} \left| \psi_{\alpha}(x) \right|^2$, with $\alpha \in \{ \uparrow, \downarrow \}$.
Here we are interested in the magnetic properties of the BEC in the absence of the impurity and therefore we have set $\psi_I(x)=0$ in Eq.~(\ref{GrossPitaevskii_3C}). 

Formally any self-consistent solution of Eq.~(\ref{General_self_consistent}) is an adequate solution of the GPE within the TF approximation.
However, herein, we are interested in the ground state of the system and as a consequence we  minimize the energy contribution stemming from $\langle \Psi_B | \hat{H}_S(x) | \Psi_B \rangle$ yielding $\langle \Psi_B | \hat{H}_S(x) | \Psi_B \rangle = -\frac{1}{2}\sqrt{\Delta^2(x)+\Omega_R^2}$ for all $x$. 
In this case, Eq.~(\ref{General_self_consistent}) and $\Delta(x) = \varepsilon_{\uparrow}(x) - \varepsilon_{\downarrow}(x)$ define two algebraic equations for $\varepsilon_{\alpha}(x)$, $\alpha \in \{ \uparrow, \downarrow \}$ the solution(s) of which determine the ground state(s) of the pseudospinor BEC. 

However, in order to identify the distinct phases of the Bose gas in a more transparent manner it is more convenient to work with quantities based on its wave function (and thus also its density).
Indeed, the choice $\langle \Psi_B | \hat{H}_S(x) | \Psi_B \rangle = -\frac{1}{2}\sqrt{\Delta^2(x)+\Omega_R^2}$ motivates us to express the wave function as 
\begin{equation}
\left( 
\begin{array}{c}
\psi_{\uparrow} (x) \\
\psi_{\downarrow} (x)
\end{array} \right) =  
\sqrt{n(x)}\left( 
\begin{array}{c}
\sin \varphi(x) \\
-\cos \varphi(x)
\end{array} \right),
\label{spin_solution}
\end{equation}
with $n(x) = \rho_{\uparrow}(x)+\rho_{\downarrow}(x)$ is the total BEC density. 
Accordingly, the phase factor 
\begin{equation}
\varphi(x) = - \frac{1}{2} \cos^{-1} \left( \frac{\Delta(x)}{\sqrt{\Delta^2(x) + \Omega_R^2}} \right).
\label{bloch_sphere_angle}
\end{equation}
Equation~(\ref{spin_solution}) expresses the wave function of the system solely in terms of the functions $n(x)$ and $\Delta(x)$. 
The render clear the physical interpretation of these functions, we note that the densities of the individual BEC components read
\begin{equation}
\rho^{(1)}_{\uparrow \atop \downarrow} (x) = \frac{n(x)}{2} \left[ 1 \mp \frac{\Delta(x)}{\sqrt{\Delta^2(x)+\Omega^2_R}} \right].
\label{densities_TF_approx}
\end{equation}
As such, $\Delta(x)$ signifies the degree of local spin-imbalance among the distinct components. In this sense $\Delta(x) \neq 0$ indicates spin-component immiscibility.

To derive the ground state solutions of the pseudospinor Bose gas we insert Eqs.~(\ref{spin_solution}), (\ref{bloch_sphere_angle}) into Eq.~(\ref{General_self_consistent}) resulting in  
the self-consistency equations 
\begin{equation}
\begin{split}
\mu_B =& 
\frac{n(x)}{2} \left(\frac{g_{\uparrow\uparrow} + g_{\downarrow\downarrow}}{2} + g_{\uparrow\downarrow} \right) \\ &- \frac{g_{\uparrow \uparrow} -g_{\downarrow\downarrow}}{4} \frac{n(x) \Delta(x)}{\sqrt{\Delta^2(x) + \Omega_R^2}} + \frac{1}{2} m_B \omega^2_B x^2 \\&-\frac{1}{2}\sqrt{\Delta^2(x)+\Omega_R^2}. \\
\Delta(x) =& 
\frac{n(x)}{2} \left( g_{\uparrow\uparrow} - g_{\downarrow\downarrow} \right) \\ &+ \left( \frac{g_{\uparrow\uparrow}+g_{\downarrow\downarrow}}{2} -g_{\uparrow\downarrow}  \right) \frac{n(x) \Delta(x)}{\sqrt{\Delta^2(x) + \Omega_R^2}}
\end{split}
\label{self_consistency_eqs}
\end{equation}
Notice here the appearance of the chemical potential, $\mu_{B}$, which is fixed by demanding $N = \int \mathrm{d}x~n(x)$, yielding a third equation for obtaining the TF profile. 
For simplicity, below, we consider a two-component system with $g_{\uparrow\uparrow}=g_{\downarrow\downarrow}=g$. 
Therefore, the solutions of Eq.~(\ref{self_consistency_eqs}) read
\begin{widetext}
\begin{equation}
\left\{
\begin{array}{l l}
\begin{array}{l}
{\displaystyle n(x) = \frac{1}{g} \left( \mu_B - \frac{1}{2} m_B \omega_B^2 x^2 \right)} \vspace{5pt}\\
{\displaystyle \Delta(x) = \pm \sqrt{(g-g_{\uparrow\downarrow})^{2} n^2(x) - \Omega^2_R}}
\end{array} & \text{for } |x| < R_{\rm in} = \Re \left(\sqrt{ \dfrac{2 [ (g_{\uparrow\downarrow}-g) \mu_B - g | \Omega_R |]}{|g_{\uparrow\downarrow}-g| m_B \omega_B^2}} \right) \vspace{10pt}\\
\begin{array}{l}
{\displaystyle n(x) = \frac{2}{g + g_{\uparrow\downarrow}} \left( \mu_B - \frac{1}{2} m_B \omega_B^2 x^2 + \frac{|\Omega_R|}{2} \right)} \vspace{5pt}\\
{\displaystyle \Delta(x) = 0}
\end{array} & \text{for } R_{\rm in} \leq |x| < R_{\rm out} = \sqrt{\dfrac{2 \mu_B  + | \Omega_R |}{m_B \omega_B^2}} \vspace{10pt}\\
\begin{array}{l}
{\displaystyle n(x) = 0} \vspace{5pt}\\
{\displaystyle \Delta(x) = 0}
\end{array} & \text{for } |x| \geq R_{\rm out}
\end{array} \right.
\label{solution_balanced_interactions}
\end{equation}
\end{widetext}
where $R_{{\rm in}}$ denotes the TF radius of the spin density imbalance, $\rho_{\uparrow}(x)-\rho_{\downarrow}(x) = -\Delta(x)/(g_{\uparrow \downarrow}-g)$, and $R_{{ \rm out}}$ is the TF radius of each component.  
Also, the chemical potential $\mu_B$ refers to the solution of the following algebraic equation
\begin{equation}
\begin{split}
N = & \frac{2}{g + g_{\uparrow \downarrow}} \bigg[ \left(2 \mu_B + | \Omega_R |\right)(R_{\rm out} - R_{\rm in})\\
&- \frac{1}{3} m_B \omega^2_B (R_{\rm out}^3 - R_{\rm in}^{3})\bigg] \\
& +\frac{1}{g} \left[ 2 \mu_B R_{\rm in} - \frac{1}{3}m_B \omega_B^2 R_{\rm in}^{3} \right]. \label{eq_chem_pot}
\end{split}
\end{equation}

The solutions provided in Eq.~(\ref{solution_balanced_interactions}) reveal that the pseudospinor BEC experiences a phase transition in terms of $\Omega_R$ and $g_{\uparrow \downarrow}$. 
For large Rabi couplings, i.e. $|\Omega_R| \geq \Omega_R^{\rm crit} = n(0)(g_{\uparrow\downarrow} -g)$, a unique solution exists (since $R_{\rm in}=0$ and the first branch does not contribute) characterized by a completely miscible phase with $\Delta (x) =0$ and therefore $\rho_{\uparrow}(x)=\rho_{\downarrow}(x)$. 
In contrast, reducing the Rabi coupling such that $|\Omega_R| < \Omega_R^{\rm crit}$ the ground state is characterized by the coexistence of two separate spatial domains in terms of $R_{\rm in}$ and $R_{\rm out}$. 
In particular, since in this case $R_{\rm in}>0$, there is an immiscible region close to the trap center, namely for $|x| \leq R_{\rm in}$.
Since the system is symmetric to spin inversions, namely it exhibits a $\mathbb{Z}_{2}$ symmetry (see discussion in Sec.~\ref{sec:spin_symmetries}) owing to the fact that $g = g_{\uparrow\uparrow} = g_{\downarrow\downarrow}$, there are two distinct solutions in this regime.
These possess either $\Delta(x)< 0$ or $\Delta(x)> 0$, corresponding to a predominant occupation of the spin-$\uparrow$ or the spin-$\downarrow$ state at the trap center respectively, see Eq.~(\ref{densities_TF_approx}). 
Therefore, we conclude that in the immiscible case the ground state of the BEC is doubly degenerate.
Furthermore, assuming a non-zero Rabi coupling, $\Omega_R \neq 0$, we obtain $R_{\rm in} < R_{\rm out}$ and therefore a miscible region always appears for $R_{\rm in} \leq | x | \leq R_{\rm out}$.
In this spatial extent the densities of the components are perfectly overlapping, as in the case $|\Omega_R| \geq \Omega_R^{\rm crit}$ discussed above. 
Notice that this {\it partially immiscible regime} does not appear for $g_{\uparrow\downarrow}< g$, since the state of the Bose gas is miscible even for $\Omega_R = 0$. Indeed, in this case it is impossible to satisfy the immiscibility condition $|\Omega_R| < \Omega_R^{\rm crit} = n(0) (g_{\uparrow\downarrow} -g)$, because $n(0) > 0$ and $|\Omega_R|>0$.

\begin{figure*}[ht]
\includegraphics[width=1.0\textwidth]{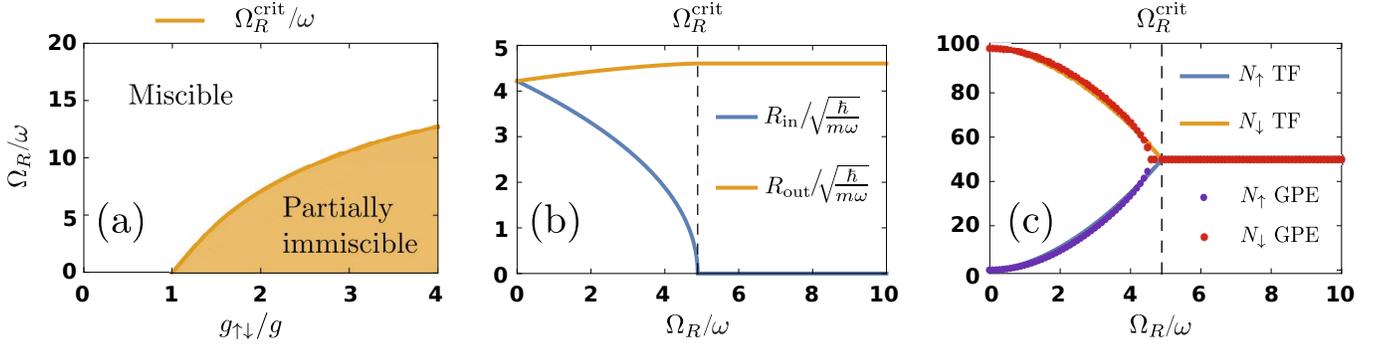}
\caption{Ground state properties of the pseudospinor BEC in the absence of impurities. (a) Phase diagram for $g= 0.5$ and $N=100$. (b) Dependence of the TF radii $R_{\rm in}$ and $R_{\rm out}$ on $\Omega_R$ for $g_{\uparrow \downarrow} = 0.8$, $g= 0.5$ and $N=100$. (c) Number of atoms in the $\ket{\uparrow}$ ($N_{\uparrow}$) and $\ket{\downarrow}$ ($N_{\downarrow}$) components for varying $\Omega_R$ within the TF and the GPE frameworks. 
The system parameters are the same as in (b). }
\label{fig:psBEC_pd} 
\end{figure*} 

To gain deeper insight into the behavior of the pseudospinor BEC for varying $\Omega_R$ and $g_{\uparrow \downarrow}$ we present its underlying phase diagram in Fig.~\ref{fig:psBEC_pd} (a) for the parameters employed in the main text, namely $g =0.5$ and $N =100$. 
Apparently, the partially immiscible phase emerges for strong $g_{\uparrow\downarrow}$ and weak $\Omega_R$. 
Furthermore, the variation of $R_{\rm in}$ and $R_{\rm out}$ for $g_{\uparrow\downarrow} = 0.8$ is illustrated in Fig.~\ref{fig:psBEC_pd} (b).
It can be readily verified that $R_{\rm in}=R_{\rm out}$ holds for $\Omega_{R}=0$, while an increasing Rabi-coupling leads to a decrease of $R_{\rm in}$, ultimately reaching $R_{\rm in}=0$ for $\Omega_{R} = \Omega_R^{\rm crit}$. 
In contrast, $R_{\rm out}$ increases compensating for the constant particle number ($N$) in the BEC. 
The miscible-immiscible phase transition for larger $\Omega_R$ is directly captured by the population of the individual component $N_{\alpha} = \int \mathrm{d}x~\rho_{\alpha}(x)$ depicted in Fig.~\ref{fig:psBEC_pd} (c).
As it can be seen within the TF approximation and for $\Omega_R = 0$ the system is fully polarized along the $z$ spin axis.
For higher values of $\Omega_{R}$ the particle imbalance decreases and eventually vanishes within the miscible regime. 

It should be, however, stressed that the TF approximation is not quantitatively reliable in the immiscible regime since the derivative of the component wave functions, $\Psi_{\alpha}(x)$, is discontinuous at $|x| = R_{\rm in}$\footnote{Notice that $\lim_{x \to R_{\rm in}^-} \frac{d}{dx} \sqrt{\rho_{\alpha}(x)}-\lim_{x \to R_{\rm in}^+} \frac{d}{dx} \sqrt{\rho_{\alpha}(x)} \neq 0$ where the appropriate for each region $n(x)$ and $\Delta(x)$ given in Eq.~(\ref{solution_balanced_interactions}) are substituted to Eq.~(\ref{densities_TF_approx}).}.
Therefore, the kinetic energy of the BEC becomes infinite in the vicinity of these points.
In order to correctly evaluate the BEC density profiles of the partially immiscible system one has to solve the full GPE equations of motion [Eq.~(\ref{GrossPitaevskii_3C})] and thus account for the impact of the kinetic energy term.
Indeed, upon considering the GPE modeled $N_{\alpha}$ is altered in the vicinity of the transition region from the TF results, see Fig.~\ref{fig:psBEC_pd} (c).
Nevertheless, the TF approximation is found to provide an adequate qualitative description of the phase of the Bose gas. 
Below, we shall employ this approximation to elucidate the phase diagram of the impurity immersed in a pseudospinor BEC.

\section{Effective potential and its range of applicability}\label{app:effective_pot}

Previous studies focussing on a spinless BEC host~\cite{mistakidis2020many,mistakidis2019quench,mistakidis2021radiofrequency,mistakidis2020pump} argued that important insights regarding the equilibrium and the dynamical Bose polaron properties are obtained by merely considering the Bose gas acting as a material barrier to the impurity. 
This is the concept of the so-called effective potential which of course neglects correlations among the impurity and the medium. 
Its success can be traced back to the separation of energy scales between density-density interactions giving rise to the effective potential and the corresponding bath-phonon interactions predicted within the Fr{\"o}hlich model~\cite{grusdt2015new,devreese2009frohlich}.
Indeed, the energy scale of density-density interactions is $E_{\rm den-den} \sim g_{BI} n_{0}$, with $g_{BI}$ being the impurity-medium effective coupling strength and $n_{0}$ denoting the BEC density.
In contrast, the impurity-phonon coupling within the Fr{\"o}hlich model is $E_{\rm imp-phon} \sim g_{BI} \sqrt{n_{0} / \xi}$~\cite{grusdt2015new}, where $\xi$ is the BEC healing length. 
Thus, the energy scale defined by the density-density interactions is larger, since for a BEC the healing length should be larger than the corresponding interparticle distance $E_{\rm den-den}/E_{\rm imp-phon} \sim \sqrt{n_0 \xi} \gg 1$~\cite{PitaevskiiStringari2016}. 
Consequently, the impact of the effective-potential is more pronounced than the phononic dressing.

The generalized form of the effective potential for the case of a pseudospinor BEC reads
\begin{equation}
V_{\rm eff}(x) = \frac{1}{2} m_I \omega^2_I x^2 + g_{I \uparrow} \rho_{\uparrow} (x)+ g_{I \downarrow} \rho_{\downarrow} (x),
\label{eff_pot_gen}
\end{equation} 
where $\rho_{\alpha}(x)$, with $\alpha \in \{ \uparrow, \downarrow \}$, is the density of the corresponding BEC component for $g_{I\uparrow} = g_{I\downarrow} = 0$.
To establish the limitations of the effective potential approach, $V_{\rm eff}(x)$ is also compared to the improved effective potential  $V'_{\rm eff}(x)$. 
The latter possesses the same form as $V_{\rm eff}(x)$ but $\rho_{\alpha}(x)$ corresponds to the BEC density calculated for the proper values of $g_{I \uparrow}$, $g_{I \downarrow}$ within ML-MCTDHX.
Therefore, $V'_{\rm eff}(x)$ incorporates effects stemming from the impurity-medium correlations allowing us to characterize their impact in the polaronic state and identify effects beyond the effective potential framework. 
This inclusion is important due to the pseudospin degree-of-freedom of the BEC.
As shown in Ref.~\cite{abad2013study} except for the phononic excitations (which retain the same structure as in the spinless scenario within the miscible regime, $\Omega_R< \Omega_R^{\rm crit}$) a new branch of spin-excitations emerges referring to out-of-phase density fluctuations among the BEC components. 
Here we are not interested in a detailed study of the coupling mechanism between spin-excitations and the impurity. 
However, such contributions lead to pronounced impurity-medium correlations beyond $V_{\rm eff}(x)$. 
These can be directly captured by comparing $V_{\rm eff}(x)$ with $V'_{\rm eff}(x)$.

Below, we derive the explicit form of $V_{\rm eff}(x)$ [Eq.~(\ref{eff_pot_gen})] within the TF approximation utilizing the results of Appendix~\ref{app:mean_field_binary}, for both miscible and immiscible BEC components.
This exploration will allow us to interpret the phase diagram of the different polaronic excitations emanating for repulsive impurity-medium interactions but also unravel the limitations of this effective approach. 

\subsection{Miscible regime}

\begin{figure*}[ht]
\includegraphics[width=1.0\textwidth]{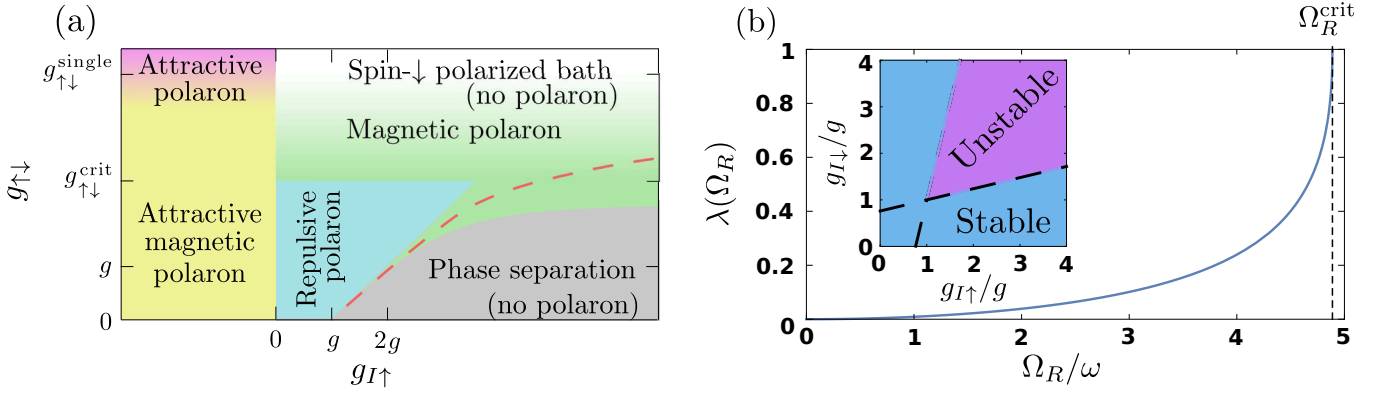}
\caption{ 
(a) Phase diagram of the Bose polaron for varying impurity-spin ($g_{I \uparrow}$, $g_{I \downarrow}=0$) and intercomponent ($g_{\uparrow \downarrow}$) interactions.
The dashed line in (a) indicates $g_{\rm mag~c}$ being the maximal value of $g_{I \uparrow}$ for which the repulsive magnetic polaron is the ground state of the system for a given $g_{\uparrow \downarrow}$.
$g_{\uparrow \downarrow}^{\rm crit}$ refers to the critical bath intercomponent interaction, for given $\Omega_R^{\rm crit}$, determining the miscible-immiscible phase-transition of the spinor Bose gas.
$g_{\uparrow \downarrow}^{\rm single}$ corresponds to the intercomponent interaction strength above which magnetic effects diminish.
(b) The factor $\lambda(\Omega_R)$ determining the stability regime of the Bose polaron in the case of an immiscible spinor BEC within the TF approximation for $g_{\uparrow \downarrow}/g=1.6$, $g = 0.5$, $N=100$, $\omega_I = \omega_B$ and $m_I = m_B$.
The inset depicts an exemplary stability diagram in the $g_{I \uparrow}$-$g_{I \downarrow}$ plane for $\Omega_R = 4 \omega$.
The dashed lines correspond to the boundaries of the stability region $g_{I \downarrow} = g + \lambda(\Omega_R = 4 \omega) (g_{I \downarrow} -g)$ and $g_{I \uparrow} = g + \lambda(\Omega_R = 4 \omega) (g_{I \uparrow} -g)$.
}
\label{fig:appendix_phase} 
\end{figure*}

Referring to miscible spinor components ($R_{\rm in}=0$), according to Eq.~(\ref{solution_balanced_interactions}), the effective potential reads
\begin{equation}
V_{\rm eff}(x) = \left\{
\begin{array}{l l}
\begin{split}
& \dfrac{1}{2} m_{B} D_{\rm misc} \omega_B^2 x^2 \\
&+ \dfrac{g_{I\uparrow} + g_{I \downarrow}}{g + g_{\uparrow\downarrow}} \left( \mu_B + \dfrac{| \Omega_R |}{2} \right),
\end{split} & \text{for } |x| \leq R_{\rm out}, \\
\dfrac{1}{2} m_I \omega_I^2 x^2, & \text{for } |x| > R_{\rm out},
\end{array} \right.
\end{equation}
here the factor $D_{\rm misc}$ determines the shape of the effective potential and reads
\begin{equation}
D_{\rm misc} = \dfrac{m_I \omega_I^2}{m_B \omega_B^2} - \dfrac{g_{I\uparrow} + g_{I \downarrow}}{g+g_{\uparrow\downarrow}}.
\end{equation}
Therefore, $V_{\rm eff} (x)$ possesses a harmonic oscillator shape for $D_{\rm misc}>0$, or equivalently $g_{I\uparrow} + g_{I\downarrow} < g_{BI {\rm crit}} =  (g + g_{\uparrow\downarrow}) \frac{m_I \omega_I^2}{m_B \omega_B^2}$ deforming into a double-well in the opposite case, $D_{\rm misc}<0$, with  minima located at $x_{\rm min} = \pm R_{\rm out}$. 
As such, in line with the results of Ref.~\cite{mistakidis2020many,mistakidis2021radiofrequency} for a single-component host, we expect that the impurity and the medium bath become immiscible in the case of 
$g_{I \uparrow} + g_{I \downarrow}>g_{BI {\rm crit}}$. 
This ultimately results in the instability of the repulsive Bose polaron being inherently related to its decaying character in this interspecies interaction regime~\cite{mistakidis2020many,mistakidis2019quench}.
This provides the first threshold for the stability of the Bose polaron indicated in Fig.~ \ref{fig:appendix_phase}(a), see in particular the edge of the blue region for $g_{\uparrow \downarrow} = g_{I \uparrow} - g$. 
This is equivalent to the edge of the repulsive Bose polaron regime for $g_{I \uparrow} = g + g_{\uparrow \downarrow}$ appearing in Fig.~\ref{fig:schematic_phase}(a).

It should be emphasized that within $V_{\rm eff}(x)$ we completely neglect backaction effects of the impurity to the BEC, which in the case of a pseudo-spinor Rabi-coupled BEC can be important for determining the stability of the Bose polaron. 
Nevertheless, by comparing $V_{\rm eff}(x)$ with the corrected $V'_{\rm eff}(x)$ potential incorporating the effect of impurity-medium correlations we do not observe significant deviations among the two approaches as long as, the Rabi-coupling is strong enough, see Fig.~\ref{fig:schematic_phase}(b$_2$) for the repulsive polaron and Fig.~\ref{fig:schematic_phase}(b$_3$) for the immiscible impurity-medium case.
Similar findings are obtained for $g_{\uparrow \downarrow} \ll g$ (not shown here for brevity). 
The above can be interpreted in view of Ref.~ \cite{abad2013study}, where it was demonstrated that the spin-excitations possess a sizable gap when the system is deep into the miscible regime of the pseudospinor BEC. 
Therefore, within the latter regime the generation of spin-wave excitations into the host is prohibited. 
Recall that also the phonon-impurity coupling is much weaker than the $V_{\rm eff}(x)$ contribution, thus suppressing possible density excitations of the medium. 
As such, both excitations pathways are essentially frozen within this miscible BEC regime and consequently the impurity-medium correlations are expected to be insignificant. 
This explains the fact that $V_{\rm eff}(x) \approx V'_{\rm eff}(x)$ as shown in Fig.~\ref{fig:schematic_phase}(b$_2$), (b$_3$). 

As the miscibility-immiscibility threshold is approached, i.e. for $\Omega_R \approx \Omega_R^{\rm crit}$ or $g_{\uparrow \downarrow} \approx g_{\uparrow \downarrow}^{\rm crit}$, the above-mentioned energy gap among the spin and the phononic excitations closes~\cite{abad2013study} allowing for the coupling of the impurity state with the spin-fluctuations of its host.
As a consequence we expect pronounced impurity-medium correlations in this regime.
Indeed, by incorporating this correction into a new effective potential, $V'_{\rm eff}(x)$, we observe the emergence of an additional well around $x=0$ which is responsible for binding the impurity into the Bose gas provided that $V'_{\rm eff}(0) < V'_{\rm eff}(R_{\rm out})$, see Fig.~\ref{fig:schematic_phase}(b$_4$). 
This behavior is reminiscent of the so-called Pekkar polaron according to which the density excitations of the host are maximized in the vicinity of the impurity leading to its spatial localization~\cite{pekar1946local,landau1948effective}. 
Importantly though, in our case the excitations dressing the impurity possess a magnetic character (see Sec.~\ref{spin_corel}) and therefore we refer to the structure emerging in this regime as the Pekkar-magnetic polaron. 

\subsection{Immiscible spin interactions}

For a medium with immiscible spin components one has to consider a larger number of cases since depending on the sign of $g_{I \uparrow} - g_{I\downarrow}$ distinct medium configurations are favored. 
Indeed, if $g_{I \uparrow} > g_{I \downarrow}$ ($g_{I \uparrow} < g_{I \downarrow}$) it is preferable that $n_{\uparrow} > n_{\downarrow}$, $\Delta(x) <0$ ($n_{\uparrow} < n_{\downarrow}$, $\Delta(x) >0$). 
This leads to a discontinuity in $n_{\alpha}$ at $g_{I \uparrow} = g_{I \downarrow}$. 
Then the effective potential in the TF approximation becomes 
\begin{widetext}
\begin{equation}
V_{\rm eff}(x) = \left\{
\begin{array}{l l}
\begin{split}
&\left( \dfrac{m_I \omega_I^2}{m_B \omega_B^2} - \dfrac{g_{I\uparrow} + g_{I \downarrow}}{2 g} \right) \dfrac{1}{2} m_{B} \omega_B^2 x^2  + \dfrac{g_{I \uparrow} + g_{I\downarrow}}{2 g} \mu_B\\
&- \dfrac{| g_{I\uparrow} - g_{I \downarrow} |}{2(g_{\uparrow\downarrow} -g)} \sqrt{ \left[ \dfrac{g_{\uparrow \downarrow} - g}{g} \left(\mu_B - \dfrac{1}{2} m_B \omega^2_B x^2 \right) \right]^2 - \Omega_R^2},
\end{split} & \text{for } |x| \leq R_{\rm in}, \\
\left( \dfrac{m_I \omega_I^2}{m_B \omega_B^2} - \dfrac{g_{I\uparrow} + g_{I \downarrow}}{g+g_{\uparrow\downarrow}}\right) \dfrac{1}{2} m_{B} \omega_B^2 x^2 + \dfrac{g_{I\uparrow} + g_{I \downarrow}}{g + g_{\uparrow\downarrow}} \left( \mu_B + \dfrac{| \Omega_R |}{2} \right), & \text{for } R_{\rm in} < |x| \leq R_{\rm out}, \\
\dfrac{1}{2} m_I \omega_I^2 x^2, & \text{for } |x| > R_{\rm out}.
\end{array}
\right. \label{eq:effective_pot_immiscible_TF}
\end{equation}
\end{widetext}

To further elucidate the form of $V_{\rm eff}(x)$ we perform an expansion with respect to $x$ around $|x| \leq R_{\rm in}$.
Within the harmonic approximation (i.e. dropping terms $\propto x^n$ for $n>2$) we obtain 
\begin{equation}
\begin{split}
&V^{\rm approx}_{\rm eff}(x) =\frac{1}{2} m_{B} D_{\rm im} \omega_B^2 x^2 \\
&+ \frac{\mu_B}{2 g} \left[g_{I\uparrow} + g_{I\downarrow} -\left| g_{I\uparrow} -g_{I\downarrow} \right| \sqrt{1 - \left( \frac{\Omega_R}{\Omega_R^{\rm crit}} \right)^2} \right] \\
&+ \mathcal{O}\left[ \left( \frac{x}{\sqrt{\mu_B/(m_B \omega_B^2)}} \right)^4 \right]
\end{split}
\end{equation}
Consequently, whether $V_{\rm eff}(x)$ exhibits a double- or single-well structure depends on the sign of 
\begin{equation}
\begin{split}
D_{\rm im} = \frac{m_I \omega_I^2}{m_B \omega_B^2} - \frac{1}{2 g} \Bigg[&g_{I \uparrow}+g_{I \downarrow} \\
&- \frac{|g_{I \uparrow} - g_{I \downarrow}|}{\sqrt{1 - \left(\frac{ \Omega_{R}}{\Omega_{R}^{\rm crit}} \right)^2}} \Bigg].  
\end{split}
\label{discriminant_immiscible}
\end{equation}
Indeed, $D_{\rm im}>0$ designates a single-well potential around $x=0$, whilst $D_{\rm im}< 0$ implies a double-well with minima at $x = \pm R_{\rm out}$. 
A first simplification occurs in the case of $g_{I \downarrow} =0$ (due to the $\mathbb{Z}_2$ symmetry we can equivalently assume $g_{I\uparrow}=0$), where $D_{\rm im}>0$ for all $g_{I \uparrow}$. This means that in the immiscible case and for $g_{I \downarrow} = 0$, the impurity can not escape from the BEC within the TF approximation and the polaron exists for every $g_{I \uparrow}$. 

By comparing, $V_{\rm eff}(x)$ with $V'_{\rm eff}(x)$ within the immiscible regime, we observe a qualitative agreement between these two approaches see Fig.~\ref{fig:schematic_phase}(b$_5$).
The notable differences lie in the modification of the effective potential around $x=0$ owing to the magnon dressing. 
This leads to a stronger binding of the impurity in the BEC when impurity-medium correlations are incorporated. 
As such, the Bose polaron in this parameter range possesses a similar structure as the magnetic one identified in the miscible regime, despite that the former does not show  a self-bound character.  
For this reason we do not differentiate among these two regimes in the phase diagram of Fig.~\ref{fig:schematic_phase}(a), (b) and refer to the emerging structures as magnetic Bose polarons. An additional alteration of $V'_{\rm eff}(x)$ occurs near $x = \pm R_{\rm in}$, originating from the non-negligible kinetic energy of the bosons in this spatial region which is neglected within the TF approximation. 
Further, for $g_{I \uparrow} = 0$ (or $g_{I \downarrow} = 0$), we note that deep in the immiscible regime, i.e. for $\Omega_R \ll \Omega_R^{\rm crit}$ (or equivalently for $g_{\uparrow \downarrow} \gg g_{\uparrow \downarrow}^{\rm crit}$) the BEC tends to be fully polarized. 
Then, all atoms occupy the spin-state that is non-interacting with the impurity and therefore in this case no polaron exists.
Of course, as the system approaches this regime it exhibits a crossover character and consequently no phase-boundary emerges.
In order to estimate the parameter region where the behavior of the system changes we define, $g_{\uparrow \downarrow}^{\rm single}$ possessing the property that for $g_{\uparrow \downarrow} > g_{\uparrow \downarrow}^{\rm single}$ and fixed $\Omega_R$ less than a single BEC atom occupies the component that is interacting with the impurity, see Fig.~\ref{fig:appendix_phase}(a).
Similarly, we have defined $\Omega_R^{\rm single}$, see Fig.~\ref{fig:schematic_phase}(a), such that for $\Omega_R < \Omega_R^{\rm single}$ and fixed $g_{\uparrow \downarrow}$ the component interacting with the impurity is occupied by less than one atom.

Let us now briefly comment on the case of $g_{I \uparrow} \neq 0 \neq g_{I \downarrow}$, by examining the expected regimes where a polaron appears within the TF approximation (i.e. $D_{\rm im} >0$). 
The behavior of the system is the simplest for $\Omega_R \ll \Omega_R^{\rm crit}$, where the spin components are strongly imbalanced and 
almost only one of them is occupied [see also Fig.~\ref{fig:psBEC_pd}]. 
It naturally follows from Eq.~(\ref{discriminant_immiscible}) that the polaron exists for $\min(g_{I\uparrow},g_{I \downarrow})< g \frac{m_I \omega_I^2}{m_B \omega^2_B}$. 
These results are in agreement to the ones regarding an impurity immersed in a spinless BEC~\cite{mistakidis2020many,mistakidis2019quench,mistakidis2021radiofrequency}.
On the contrary, in the region $\Omega_R \approx \Omega_R^{\rm crit}$, i.e. close to the miscibility-immiscibility threshold, $D_{\rm imm}<0$ signifying that a potential minimum at $x = 0$ always exists independently of the impurity-medium interaction strength.
Finally, in the intermediate case, $0 < \Omega_R <\Omega_R^{crit}$, there is a threshold 
\begin{equation}
\begin{split}
\min(g_{I \uparrow},g_{I \downarrow}) &- g \frac{m_I \omega^2_I}{m_B \omega_B^2} < \lambda(\Omega_{R}) \times \\
&\left( \max(g_{I \uparrow},g_{I \downarrow}) - g \frac{m_I \omega^2_I}{m_B \omega_B^2} \right),
\end{split}
\end{equation}
below which the Bose polaron exists. 
Here, the function $\lambda(\Omega_R)$ features an increasing behavior for $0 < \Omega_R <\Omega_R^{crit}$, and exhibits the extrema $\lambda(\Omega_R=0)=0$ and $\lambda(\Omega_R=\Omega_R^{\rm crit})=1$. 
The precise values of $\lambda(\Omega_R)$ in the above-mentioned domain, depend solely on the parameters characterizing the medium, namely $g$, $g_{\uparrow \downarrow}$ and $N$. 
For illustration, Fig.~\ref{fig:appendix_phase}(b) provides $\lambda(\Omega_R)$ alongside with an example stability phase diagram in the inset for the host parameters employed within this work.

Concluding, $V_{\rm eff}(x)$ provides insight into the existence of polaronic states, but not whether they correspond to the system's ground state. 
Indeed, if the impurity remains within the BEC its energy increases without bound for a larger $g_{I \alpha}$.
In contrast, the phase-separated states possess an energy that saturates for strongly repulsive impurity-medium interaction strengths since the interspecies spatial overlap is small. 
Therefore, it is expected that the impurity energy in these cases will saturate towards $E_{\rm ph. s.} \sim \frac{1}{2} m_I \omega_I^2 R_{\rm out}^2$. 
This implies that even in the cases that the Bose polaron persists for large interactions, the phase-separated states might be the overall ground states of the system and therefore explicit investigations are required to determine the energy of both branches. 
Here a hint towards the stability of the polaronic states relies on the fact that the state of the BEC is highly perturbed for the cases of the magnetic polaron and almost completely intact for a phase-separated impurity-medium configuration. 
As a consequence, the corresponding coupling among the polaronic and phase-separated states that would render the former configuration unstable is small. 
This can be understood from the fact that the state overlap typically scales as $\propto [\sum_{\alpha} \int {\rm d}x~\sqrt{\rho_{\alpha}(x) \rho'_{\alpha}(x)}]^N$, where $\rho_{\alpha}(x)$, $\rho'_{\alpha}(x)$ are the original and perturbed single-particle densities respectively.

\section{Variational treatment of the magnetic polaron} \label{app:variational_treat} 

In order to capture the correlation properties of the magnetic polaron we solve the many-body Schr{\"o}dinger equation of the underlying multicomponent system using the variational ML-MCTDHX method~\cite{cao2017unified}. 
Specifically, a two-step truncation scheme is performed in the many-body wave function. 
First, the impurity-medium correlations are taken into account by expanding the many-body wave function  $|\Psi(t)\rangle$ with respect to $D$ orthonormal species functions, i.e.  
$|\Psi^{\sigma}_k(t)\rangle$ with  $k=1,2,\dots,D$~\cite{cao2017unified} for each component $\sigma=B,I$.  
Namely, we use the following truncated Schmidt decomposition      
\begin{equation} 
|\Psi(t)\rangle=\sum_{k=1}^D \sqrt{\lambda_k(t)} |\Psi^{\rm B}_k(t)\rangle|\Psi^{\rm I}_k(t)\rangle,  
\label{eq:wfn}
\end{equation} 
with the expansion coefficients $\lambda_k$ known as the Schmidt weights and corresponding to the eigenvalues of the $\sigma$-component reduced density matrix. 
The latter is 
$\rho_{\sigma}^{N_{\sigma}} (\vec{x}, \vec{x}';t)= \braket{\Psi(t)|\prod_{i=1}^{N_{\sigma}} \Psi_{\sigma}^{\dagger} 
(x_i) \prod_{i=1}^{N_{\sigma}} \Psi_{\sigma} 
(x_i')|\Psi(t)} $, with $\vec{x}=(x_1, \cdots,x_{N_{\sigma}})$. 
A pre-requisite for the system to be entangled~\cite{horodecki2009quantum}, or otherwise impurity-medium correlations are present, is that at least two different $\lambda_k$ are populated.

Next, we express each of the above-described species functions as a linear superposition of time-dependent number-states $|\vec{n} (t) \rangle^{\sigma}$ with time-dependent expansion coefficients, $A^{\sigma}_{i;\vec{n}}(t)$, 
\begin{equation}
    | \Psi_i^{\sigma} (t) \rangle =\sum_{\vec{n}} A^{\sigma}_{i;\vec{n}}(t) | \vec{n} (t) \rangle_{\sigma}.  
    \label{eq:number_states}
\end{equation} 
A particular number state $|\vec{n} (t) \rangle_{\sigma}\equiv \ket{n_1,\dots,n_{d_{\sigma}}}$ corresponds to a permanent. It is constructed by $d_{\sigma}$ time-dependent variationally optimized single-particle functions (SPFs) $\left|\phi_l^{\sigma} (t) \right\rangle$, where $l=1,2,\dots,d_{\sigma}$ and $n_l$ denote their occupation numbers. At this stage of the wave function truncation we account for intracomponent correlations. 

Finally, the SPFs are expanded on a time-independent primitive basis. 
The latter is an $\mathcal{M}$ dimensional discrete variable representation (DVR) for the impurity denoted by $\lbrace \left| k \right\rangle \rbrace$. For the spinor bosonic medium it is the tensor product ($\lbrace \left| k,s \right\rangle \rbrace$,) of the DVR basis for the spatial degrees-of-freedom and the two-dimensional pseudospin-$1/2$ basis $\{\ket{\uparrow}, \ket{\downarrow}\}$. 
Particularly, a SPF of the medium takes the spinor wave function form  
\begin{equation}
| \phi^{\rm B}_j (t) \rangle= \sum_{k=1}^{\mathcal{M}}\big( B^{{\rm
B}}_{jk \uparrow}(t) \ket{k} \ket{\uparrow}+B^{{\rm B}}_{jk \downarrow}(t) \ket{k} \ket{\downarrow}\big),  \label{eq:spfs}
\end{equation}
with $B^{{\rm I}}_{j k \uparrow}(t)$ and $B^{{\rm I}}_{j k \downarrow}(t)$ being the time-dependent expansion coefficients of the spin-$\uparrow$, $\downarrow$ respectively, see also 
Refs.~\cite{mistakidis2020pump,mistakidis2021radiofrequency} for further details. 

To calculate the ground state of the underlying ($N_{\uparrow}+N_{\downarrow}+1$)-body wave function $\left|\Psi(t) \right\rangle$ describing the Hamiltonian of Eq.~(\ref{Hamiltonian_total}) we solve the respective ML-MCTDHX equations of motion~\cite{cao2017unified} within the imaginary time propagation method.
These equations are found upon applying, for instance, the Dirac-Frenkel variational principle~\cite{dirac1930note} for the ansatz of 
Eqs.~(\ref{eq:wfn}), (\ref{eq:number_states}) and (\ref{eq:spfs}). 
This process leads to a coupled set of $D^2$ linear differential equations 
of motion for the $\lambda_k(t)$ coefficients as well as $D(\frac{(N+d_B-1)!}{N!(d_B-1)!}+\frac{(N_I+d_I-1)!}{N_I!(d_I-1)!})$ and $d_B+d_I$ nonlinear integrodifferential 
equations for the species functions and the SPFs respectively. 
The orbital configuration space $C=(D;d_B;d_I)$ assigns the Hilbert space truncation.  
Here, the weakly interacting spinor Bose gas has a mesoscopic atom number resulting in suppressed intracomponent correlations which can be captured by a relatively small number of orbitals, here $d_B<4$. 
For the impurity we are able to employ a substantially larger orbital number, herein $d_I=8$, in order to describe strong impurity-medium correlations. 
As such, it is feasible to numerically solve the ML-MCTDHX equations of motion.

\bibliographystyle{apsrev4-1}
\bibliography{reference}	

\end{document}